\documentclass[%
 reprint,
superscriptaddress,
nofootinbib,
 amsmath,amssymb,
 aps,
showkeys
]{revtex4-2}

\UseRawInputEncoding

\usepackage{graphicx}
\graphicspath{ {./images/} }
\usepackage{dcolumn}
\usepackage{bm}
\usepackage{hyperref}
\usepackage{xcolor}

\hypersetup{
    colorlinks=true,
    linkcolor=blue,
    filecolor=magenta,      
    urlcolor=blue,
}

\newcommand{\Rvec}{\mathbf{\hat{R}}}
\newcommand{\Svec}{\mathbf{\hat{S}}}

\begin{document}

\preprint{APS/123-QED}

\title{FastEMRIWaveforms: New tools for millihertz gravitational-wave data analysis}

\author{Michael L. Katz}
\affiliation{Max-Planck-Institut f\"ur Gravitationsphysik, Albert-Einstein-Institut, 
Am M\"uhlenberg 1, 14476 Potsdam-Golm, Germany}
\email{michael.katz@aei.mpg.de}
 
\author{Alvin J. K. Chua}
\affiliation{
Theoretical Astrophysics Group, California Institute of Technology, Pasadena, CA 91125, United States
}

\author{Lorenzo Speri}
 \affiliation{Max-Planck-Institut f\"ur Gravitationsphysik, Albert-Einstein-Institut,
Am M\"uhlenberg 1, 14476 Potsdam-Golm, Germany}

\author{Niels Warburton}
\affiliation{%
School of Mathematics and Statistics, University College Dublin, Belfield, Dublin 4, Ireland
}

\author{Scott A. Hughes}
\affiliation{%
Department of Physics and MIT Kavli Institute,\\
Massachusetts Institute of Technology, Cambridge, MA 02139, USA
}

\begin{abstract}
We present the \texttt{FastEMRIWaveforms} (FEW) package, a collection of tools to build and analyze  extreme mass ratio inspiral (EMRI) waveforms. Here, we expand on the {\it Physical Review Letter} that introduced the first fast and accurate fully-relativistic EMRI waveform template model. We discuss the construction of the overall framework; constituent modules; and the general methods used to accelerate EMRI waveforms. Because the fully relativistic FEW model waveforms are for now limited to eccentric orbits in the Schwarzschild spacetime, we also introduce an improved Augmented Analytic Kludge (AAK) model that describes generic Kerr inspirals. Both waveform models can be accelerated using graphics processing unit (GPU) hardware. With the GPU-accelerated waveforms in hand, a variety of studies are performed including an analysis of EMRI mode content, template mismatch, and fully Bayesian Markov Chain Monte Carlo-based EMRI parameter estimation. We find relativistic EMRI waveform templates can be generated with fewer harmonic modes ($\sim10-100$) without biasing signal extraction. However, we show for the first time that extraction of a relativistic injection with semi-relativistic amplitudes can lead to strong bias and anomalous structure in the posterior distribution for certain regions of parameter space.


\end{abstract}

\keywords{gravitational waves, extreme mass ratio inspirals, LISA, computational methods}
      
\maketitle


\section{Introduction}\label{sec:intro}

Gravitational wave observations from ground-based detectors are providing many new insights into the relativistic universe \cite{O2_summary, LVK2020_prospects}. The future space-based Laser Interferometer Space Antenna (LISA) will complement this science by enabling observations in the milliHertz regime \cite{LISAMissionProposal}. This region of the spectrum is rich in sources including Galactic double white dwarf (WD) binaries, massive black hole binaries (MBH), and stellar origin black hole binaries (SOBH) early in their evolution. Another key class of sources are extreme mass-ratio inspirals (EMRIs). These are compact binaries with a mass ratio $\mu/M\simeq 10^{-4}-10^{-7}$ where $\mu\sim1-100M_\odot$ is the mass of the orbiting secondary and $M\sim10^{5}-10^{7}M_\odot$ is the mass of the MBH. EMRIs are expected to form in dense stellar clusters of galactic nuclei \cite{AmaroSeoane:2012tx, Pan2021EMRIFormation} where their formation rate ranges from $\sim1-10^4$ per year with observable signal-to-noise ratios (SNR, $\rho$) expected to be $\sim20-100$ over the duration of the signal \cite{Babak2017, Amaro-Seoane2007, Porter2009LISADataOverview}. The details depend on the precise formation mechanism but it is anticipated that the majority of EMRIs will be highly eccentric, precessing binaries \cite{Babak2017}. This means that EMRIs have some of the richest and most complicated gravitational waveforms of any compact binary system. The small mass ratio of EMRIs also means that they evolve slowly and they typically have $\sim10^4-10^5$ orbits over a period of years whilst in the LISA band. The long lasting, complex waveforms of EMRIs presents a substantial challenge for both the modelling of these binaries and the LISA data analysis task.

The rewards for modelling and extracting the EMRI signals from the LISA data stream are high. The properties of the binary can be determined to sub-percent level \cite{Babak2017} which enables precision tests of general relativity in the strong field regime \cite{Barack2007, Gair2013}. Measuring EMRI parameters will also inform our understanding of the MBH mass function \cite{Gair2010}, dense stellar environment in galactic cores \cite{Amaro-Seoane2007}, and gas disks around massive black holes \cite{Barausse2007, Barausse2008, Gair2011, Yunes2011, Barausse2014, Barausse2015}.  Gravitational waves from EMRIs will also help constrain cosmological parameters \cite{MacLeod:2007jd}, including the dark energy equation of state \cite{Laghi2021EMRICosmology}.

Extracting this wealth of information from the LISA data stream will be a challenging task for two key reasons: (i) we require the waveform templates to have a phase error $\Delta \Phi \lesssim 1/\rho$; this can be as small as $\Delta \Phi \lesssim 1/100$ for a loud EMRI \cite{Amaro-Seoane2011CapraEMRIs}, and (ii) in order to search across the large parameter space we need waveforms that can be generated in less than a second. These two requirements have led to the development of two classes of EMRI models: gravitational self-force models for accuracy and ``kludge'' models for speed.

Gravitational self-force models employ black hole perturbation theory. In this approach the metric of the binary is expanded in powers of the mass ratio around the metric of the MBH.  It is known via a two-timescale analysis that this expansion must be carried out to second order (in the mass ratio) in order to meet the sub-radian accuracy goal in the GW phase \cite{Hinderer2008twotimescale}. The GW amplitudes on the other hand only need to be known to first order \cite{Hinderer2008twotimescale}. The first-order gravitational self-force is now known for generic orbits about a Kerr black hole \cite{van_de_Meent_2018} and significant progress is being made with second-order calculations \cite{Pound20202ndOrder,  Warburton:2021kwk}.

To date, development of the gravitational self-force approach has focused on calculating the inspiral motion of the secondary. The numerical computation of the gravitational self-force is slow but can be precomputed and interpolated in an offline step before the inspiral is rapidly generated (after appropriately averaging terms that oscillate on the orbital timescale if needed \cite{van_de_Meent_2018b,Miller2020_twotimescale}). Once the inspiral is known, the associated waveform can be computed. These waveform calculations tend to take tens of minutes to hours depending on the computational approach \cite{Hughes2021EMRI_FD, Warburton:2017sxk,Sundararajan:2008zm}.

Contrasting with the slow gravitational self-force waveform models are the fast EMRI ``kludge'' models which are designed to be rapidly evaluated for use in LISA data analysis studies \cite{BarackCutler2004, Babak2007, Chua2015, Chua2017}. These models capture the phenomenology of generic EMRI waveforms but only have an approximation to the correct phasing and amplitudes. They achieve this by computing the inspiral using (post-Newtonian inspired) analytic fits to pieces of the gravitational self-force and then approximating the waveform using a ``semi-relativistic'' quadrupole formula (possibly with octupolar corrections \cite{Babak2007}). This weak-field approximation fundamentally limits the improvements that can be made to these models.  Nonetheless, kludge models are currently the only EMRI templates available for use in data analysis studies that encompass the full 14-dimensional parameter space of EMRIs (neglecting the spin of the secondary). For this reason they are the only EMRI models to have been used so far in LISA data analysis studies \cite{Babak_2009, Cornish_2011}.

For future LISA data analysis, we need to combine the speed of kludge models with the accuracy of gravitational self-force models. One waveform acceleration technique that has been highly successful for comparable-mass compact binaries are reduced-order-model (ROM) surrogates. Recently this approach has been pursued in the small mass-ratio context \cite{Rifat:2019ltp} but it is not clear if these models will scale to the long signal duration of the EMRI problem.

In this work, building upon our recent {\it Letter} \cite{Chua2020RapidGenLetter}, we present the FastEMRIWaveforms (FEW) computational framework.  This framework allows gravitational self-force-based waveform models to be computed about as rapidly as kludge models, whils retaining their inherent accuracy. The key observation in our approach is that only the waveform phase needs to be known to very high precision; the amplitudes of the thousands of harmonic modes in an EMRI can computed to a much lower accuracy. Through a combination of ROM and deep-learning techniques, we produce a sufficiently accurate global fit for these amplitudes that returns the full set of modes simultaneously at each sampling time. As our mode-amplitude model is composed of simple linear-algebra operations its implementation is highly parallelizable on graphics processing units (GPUs). As a result we are able to construct the first fully relativistic, analysis length EMRI waveforms in less than 500 milliseconds. 

To showcase the FEW framework we build a module that can compute EMRI waveforms for eccentric inspirals into a non-rotating black hole. In calculating the phasing for these inspirals we use orbit-averaged pieces of the first-order gravitational self-force which produces an ``adiabatic inspiral'' \cite{Hinderer2008twotimescale,Hughes2021EMRI_FD}. These inspirals will dephase by tens to hundreds of radians over a radiation reaction timescale with respect to the true waveform \cite{Osburn:2015duj}.  The FEW framework is designed so that many of the most important higher-order gravitational self-force corrections can easily be incorporated as they become available; some effects, such as the impact of resonances \cite{speri2021assessing} (which will make relatively large contributions to EMRI phase evolution at a small number of moments during an inspiral), may be more challenging to include. The adiabatic inspirals we use in this analysis may nonetheless be useful for searches for loud EMRIs \cite{Gair2004EMRIsPreds, Chua2017}, or for binaries with a mass ratio with $\mu/M<10^{-7}$ for which the LISA mission duration is less than a radiation reaction timescale \cite{Gourgoulhon2019EMRIs_GC, Amaro-Seoane2019_XMRIs}.

Our first fully relativistic EMRI waveform model is for non-rotating black holes but our modular computational framework is set up for generic inspirals. To showcase this we add the Augmented Analytic Kludge (AAK) to the framework with an updated 5PN inspiral model and GPU acceleration. Both the relativistic and kludge models can be computed in the detector frame and are accessed via a common application programming interface (API) which streamlines interaction with LISA data analysis software. All the code is publicly available as open-source software \cite{BHPToolkit-FEW}. 

Finally, with our new model we provide the first Bayesian posterior analysis with LISA on relativistic EMRI waveforms. From this we discover that only a relatively small number of harmonic modes are required to faithfully represent the waveform. By reducing the number of harmonic modes our waveform model runtime decreases to tens of milliseconds on a GPU. We also take the opportunity to make the first study of the biases introduced in kludge models by their use of semi-relativistic waveform generation. To do this we use the modular framework of FEW to drive the AAK model with an adiabatic inspiral and compare the associated semi-relativistic waveforms with the fully relativistic one. We find that the AAK waveform generation results in significant biases in some parts of the parameter space.

The rest of this paper is structured as follows. We begin with a brief description of and equations related to building an EMRI waveform in the detector frame in Section\ref{sec:emri_wave}. In Section\ref{sec:FEW_framework} we discuss the overall framework of FEW. In Section\ref{sec:schwarz_ecc}, the first example of a relativistic waveform built within the FEW framework is described. In Section\ref{sec:newAAK}, a new and improved version of the Augmented Analytic Kludge \cite{Chua2015, Chua2017} is presented using an improved trajectory model. This waveform is also implemented in the FEW framework. In Section\ref{sec:analysis}, we analyze the mismatch, mode content, and generation of these new waveforms. We also provide the first Bayesian posterior analysis with relativistic EMRI waveforms. To conclude we discuss the future direction of the FEW framework and these models as we move towards the LISA mission in Section\ref{sec:discuss}. In this article we will geometrized units with $G=c=1$.

\section{EMRI Waveform Overview}\label{sec:emri_wave}

EMRI waveforms are represented by the complex time-domain dimensionless strain $h(t) = h_+ - ih_\times$, where $h_+$ and $h_\times$ are the normal transverse-traceless gravitational wave polarizations. At a large distance from the source, $h$ is given by \cite{DrascoHughes2006}
\begin{equation}\label{eq:main_wave}
    h = \frac{\mu}{d_L}\sum_{lmkn} A_{lmkn}(t)S_{lmkn}(t, \theta)e^{im\phi}e^{-i\Phi_{mkn}(t)},
\end{equation}
where $t$ is the time of arrival of the gravitational wave at the solar system baricenter, $\theta$ is the source-frame polar viewing angle, $\phi$ is the source-frame azimuthal viewing angle, $d_L$ is the luminosity distance, and $\{l,m,k,n\}$ are the indices describing the frequency-domain harmonic mode decomposition. The indices $l$, $m$, $k$, and $n$ label the orbital angular momentum, azimuthal, polar, and radial modes, respectively. $\Phi_{mkn}=m\Phi_\varphi + k \Phi_\theta + n\Phi_r$ is the summation of decomposed phases for each given mode. The amplitude $A_{lmkn}$ is related to the amplitude $Z^\infty_{lmkn}$ of the Teukolsky mode amplitude far from the source.  It is given by $A_{lmkn} = -2Z^\infty_{lmkn}/\omega_{mkn}^2$, where $\omega_{mkn}=m\Omega_\varphi + k \Omega_\theta + n \Omega_r$ is the frequency of the mode, and $\Omega_{r,\theta,\phi}$ describe the frequencies of a Kerr geodesic orbit.  These frequencies are determined from \cite{Fujita2009FundFreqs} using the dimensionless spin of the MBH, $a$, and the quasi-Keplerian orbital parameters of $p$ (semi-latus rectum; hereafter ``separation''), $e$ (eccentricity), and $\cos{I} \equiv x_I$ (cosine of the angle $I$ which describes the orbit's inclination from the equatorial plane).  See Ref.\ \cite{Hughes2021EMRI_FD} for further discussion and more detailed definitions.  The phases $\Phi_{\varphi,\theta,r}$ given above are the integral over time of the orbit's fundamental frequencies: 
\begin{equation}
    \Phi_{\varphi,\theta,r}=\int_0^t dt'\, \Omega_{\varphi, \theta, r}(p(t'), e(t'), x_I(t')). 
\end{equation}
The function $S_{lmkn}(t,\theta)$ is a spin-weighted spheroidal harmonic. Because these harmonics depend on the orbital frequencies, and the frequencies evolve with time, the resulting harmonics evolve as well \cite{DrascoHughes2006}. 

For a geodesic orbit, $A_{lmkn}$ and $S_{lmkn}$ are determined by solving the Teukolsky equation \cite{Teukolsky1973}. The wave field $h$ is related to the Weyl curvature scalar, $\psi_4$, at null infinity by its second derivative: $\ddot{h}=2\psi_4$. The curvature scalar can be determined for each unique set of geodesic parameters $\{a, p, e, x_I\}$ by decomposing in the frequency domain and separating the radial and angular dependencies. In this representation, $\psi_4$ is given by \cite{Teukolsky1973}
\begin{equation}
    \psi_4=\sum_{lmkn}R_{lmkn}(r)S_{lmkn}(t, \theta)e^{im\phi}e^{-i\omega_{mkn}t}.  
\end{equation}
As $r$ approaches the event horizon and infinity, the radial solution $R_{lmkn}(r)$ takes a simple limiting form, with $R_{lmkn} \propto Z^{H,\infty}_{lmkn}$.  It can then be shown that the complex amplitudes $Z^{H,\infty}_{lmkn}$ encode the rate at which an orbit evolves due to the orbit-averaged dissipative backreaction of gravitational-wave emission; the amplitude $Z^\infty_{lmkn}$, in addition,3 describes the contribution to the harmonic $lmkn$ of the gravitational waveform.  These quantities are computed for specified geodesics, and are then interpolated from geodesic to geodesic and combined with an accumulated phase to build a full waveform.  More information and details can be found in Refs.\ \cite{DrascoHughes2006, Hughes2021EMRI_FD}.

\subsection{Generating Detector-Frame Waveforms for Data Analysis}\label{sec:detector_frame}

\begin{figure*}
\begin{tabular}{ccc}
  a)\includegraphics[scale=0.8]{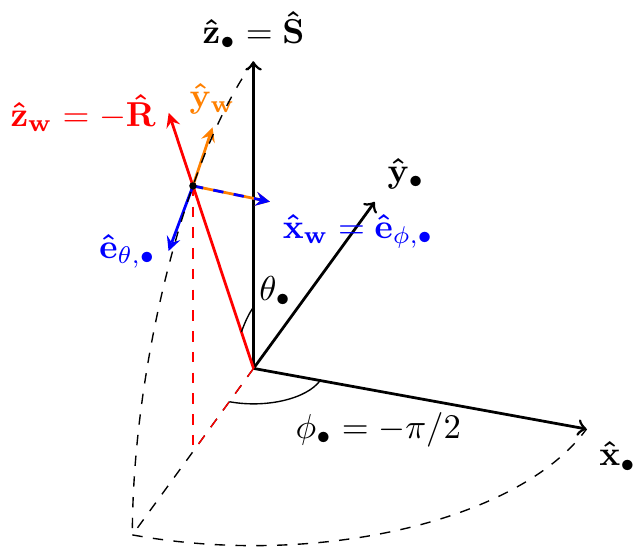} &   b)\includegraphics[scale=0.8]{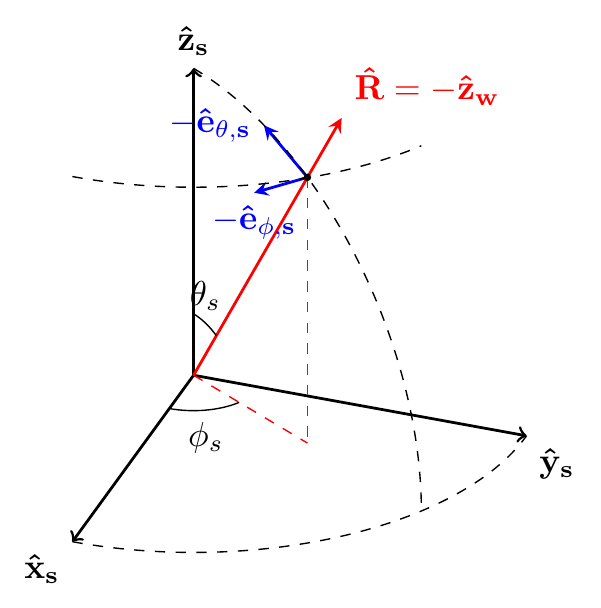} &  c)\includegraphics[scale=0.9]{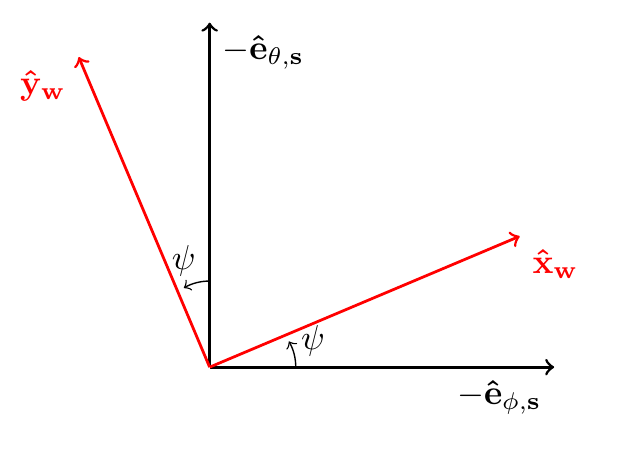}
\end{tabular}
\caption{Conventions for the source frame, solar system baricenter frame, and polarization angle are shown in the schematic diagram above. For the source-frame diagram (a),  the coordinate basis is defined in Eq.~\eqref{eq:sourceframe}. The direction towards the solar system baricenter $(-\mathbf{\hat{R}})$ is shown as the red arrow which defines $\mathbf{\hat{z}}_w$. The source-frame polarization vectors are shown in blue and the basis vectors in the wave frame (see Eq.~\ref{eq:waveframe}) are shown in orange. Note the rotation by $\pi/2$ to align the source-frame polarization vectors with the wave-frame basis. Also, due to conventions chosen, $\phi_\bullet=-\pi/2$ always. In the solar system baricenter frame (b), the solar system baricenter frame polarization basis is shown in blue. The red arrow shows the angle to the source ($+\mathbf{\hat{R}}$). Finally, in plot (c), the polarization determination is illustrated. The polarization angle represents the rotation down the line-of-sight from the wave-frame (red) to the solar system baricenter polarization basis (black).}\label{fig:conventions}
\end{figure*}

The full EMRI parameter space for a detector-frame EMRI waveform is 17-dimensional: $\{M$, $\mu$, $a$, $\vec{a}_2$, $p_0$, $e_0$, $x_{I,0}$, $d_L$, $\theta_S$, $\phi_S$, $\theta_K$, $\phi_K$, $\Phi_{\varphi,0}$, $\Phi_{\theta, 0}$, $\Phi_{r, 0}\}$. The parameters $\{M$, $\mu$, $a$, $p_0$, $e_0$, $x_{I,0}$, $d_L$, $\Phi_{\varphi,0}$, $\Phi_{\theta, 0}$, $\Phi_{r, 0}\}$ are defined above; $\theta_S$ and $\phi_S$ are the polar and azimuthal sky location angles given in the solar system barycenter frame; $\theta_K$ and $\phi_K$ are the azimuthal and polar angles describing the orientation of the spin-angular momentum vector of the MBH, $\mathbf{\hat{S}}$; and $\vec{a}_2$ is the three-dimensional spin angular momentum vector of the CO. $\vec{a}_2$ is currently ignored in waveform generation, but will be necessary for full waveform descriptions \cite{Huerta2011SpinEffects}.

To introduce detector-frame waveforms, we follow the constructions and bases used for the kludge models \cite{Chua2017}. Diagrams containing source frame, solar system baricenter frame, and polarization conventions are shown in \mbox{Figure \ref{fig:conventions}}. The sky-position vector along the line-of-sight to the EMRI system, $\hat{R}$, is given by,
\begin{equation}
    \Rvec = (\sin{\theta_S}\cos{\phi_S}, \sin{\theta_S}\sin{\phi_S}, \cos{\theta_S}),
\end{equation}
where the three components are with respect to the $(\mathbf{\hat{x}_s}, \mathbf{\hat{y}_s}, \mathbf{\hat{z}_s})$ unit vectors defined by the solar system baricenter coordinate frame basis. Similarly, $\mathbf{\hat{S}}$ in the solar system baricenter coordinate frame is given by,
\begin{equation}
    \Svec = (\sin{\theta_K}\cos{\phi_K}, \sin{\theta_K}\sin{\phi_K}, \cos{\theta_K}).
\end{equation}
When the MBH has no spin ($a=0$), $\Svec$ is equivalent to the orbital angular momentum unit vector, $\mathbf{\hat{L}}$. The wave-frame basis (denoted by a $w$ subscript) is given by \cite{Chua2017}
\begin{equation}\label{eq:waveframe}
    (\mathbf{\hat{x}}_w, \mathbf{\hat{y}}_w, \mathbf{\hat{z}}_w) := \left(\frac{\Rvec\times\Svec}{\mathcal{S}_{\Rvec, \Svec}}, \frac{\Svec - (\Svec \cdot \Rvec)\Rvec}{\mathcal{S}_{\Rvec, \Svec}}, -\Rvec \right),
\end{equation}
where $\mathbf{\hat{z}}_w$ points along the propagation direction of the wave from the source to the detector and \mbox{$\mathcal{S}_{\mathbf{\hat{S}},\mathbf{\hat{R}}} := (1-(\mathbf{\hat{S}}\cdot\mathbf{\hat{R}})^{2})^{1/2}$}. Within the kludge model framework, $\mathbf{\hat{x}}_w$ and $\mathbf{\hat{y}}_w$ represent the polarization unit vectors used to determine the $h_+$ and $h_\times$ contributions. 

For the relativistic waveforms, the source-frame (denoted with $\bullet$) viewing angles, $\theta_\bullet$ and $\phi_\bullet$ (equivalent to $\theta$ and $\phi$ in Eq.~\ref{eq:main_wave}), are determined by projecting $\Rvec$ into the source frame. We define the source frame in the same way as the kludge models:
\begin{equation}\label{eq:sourceframe}
    (\mathbf{\hat{x}}_\bullet, \mathbf{\hat{y}}_\bullet, \mathbf{\hat{z}}_\bullet) := \left(\frac{\Rvec\times\Svec}{\mathcal{S}_{\Rvec, \Svec}}, \frac{\Rvec - (\Rvec \cdot \Svec)\Svec}{\mathcal{S}_{\Rvec, \Svec}}, \Svec \right),
\end{equation}
where $\mathbf{\hat{z}}_\bullet$ points along the spin axis of the MBH and $\mathbf{\hat{x}}_\bullet$ aligns with $\mathbf{\hat{x}}_w$. This convention leads to $\phi_\bullet=-\pi/2$ always. With $\phi_\bullet$ fixed, the phasing of the wave is determined with initial phases $\{\Phi_{\varphi,0}, \Phi_{\theta, 0}, \Phi_{r, 0}\}$. The polar viewing angle is then determined by $\cos\theta_\bullet=-\Rvec\cdot\Svec$. 

In the source frame, the polarization vectors are given by $\mathbf{\hat{e}}_{\theta,\bullet}$ and $\mathbf{\hat{e}}_{\phi,\bullet}$, the conventional spherical coordinate unit vectors. To align the polarization angles of the relativistic source frame with the kludge models, a rotation angle of $\pi/2$ is applied. This rotation is equivalent to a phase shift of $\pi$.

Following this procedure, both relativistic and kludge waveforms are aligned and require a transformation to the solar ssolar system baricenterystem baricenter frame. The rotation to the solar ssolar system baricenterystem baricenter frame is twice the polarization angle, $\psi$, given by,
\begin{equation}
    \psi = \frac{\cos\theta_S\sin\theta_K\cos(\phi_S - \phi_K) - \sin\theta_S\cos\theta_K}{\sin\theta_K\sin(\phi_S - \phi_K)}.
\end{equation}

\section{Fast EMRI Waveforms Framework}\label{sec:FEW_framework}

The FEW framework was created to generate fast and accurate EMRI waveforms in the form of  Eq.~\eqref{eq:main_wave} with high fidelity when compared to slow waveform calculations. The framework was designed with specific goals in mind:
\begin{itemize}
    \item Fast waveforms must maintain a sufficiently high fidelity to accurate waveforms so as not to bias source search and parameter estimation.
    \item It must be modular so that pieces of code can be used as stand alone tools, as well as combined in a variety of ways to minimize extraneous source code.
    \item A high level of flexibility is needed so that as new physics are added to EMRI waveforms or new computational methods are devised, FEW can readily adapt and accept these changes.
    \item The user interface must be simple at all levels so that it is easy to use and contribute to.
    \item Underneath the user interface, acceleration and paralellization techniques must be utilized to make FEW waveforms as fast as possible. 
\end{itemize}

\begin{figure*}
\begin{center}
\includegraphics[scale=0.7]{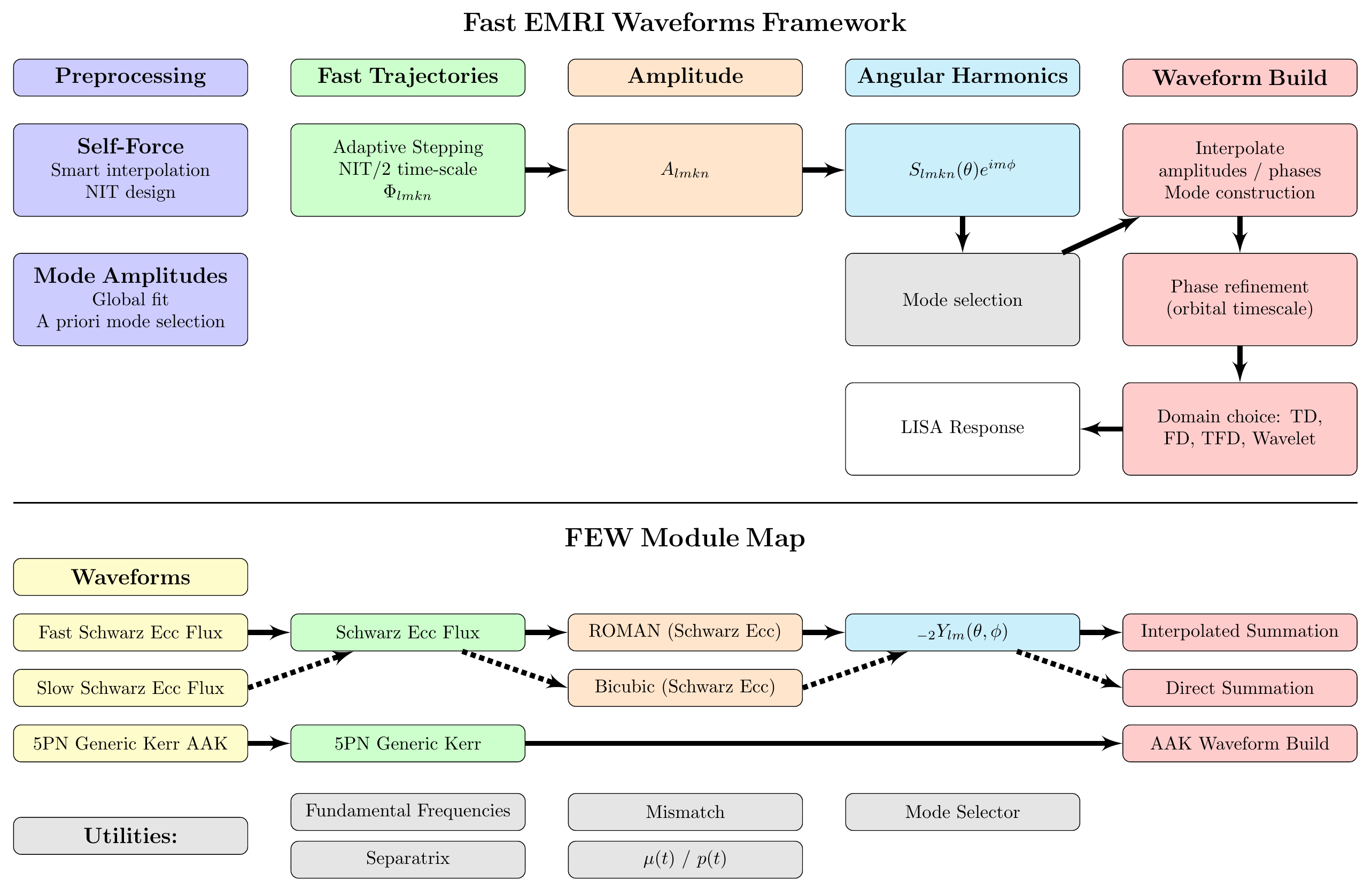}
\caption{The overall modular framework for FEW is shown in the top schematic diagram. This diagram describes the high-level progression of module usage during waveform production. Each segment of waveform production is labelled at the top of each column and assigned a color. The bottom diagram shows the waveforms and modules that have been implemented to-date in the FEW package. Stock waveforms are shown in yellow in the first column. Their individual module progression towards the final waveform is shown with arrows. The modules are assigned the color of the segment of waveform production they belong to. The utility functions that are used both for waveform production and overall analysis are also shown towards the bottom in gray.}\label{fig:emri_diagram}
\end{center}
\end{figure*}

The general FEW framework is visualized in \mbox{Figure \ref{fig:emri_diagram}}. In this section, we will discuss the base modules necessary to build the FEW framework. Each module that has GPU capabilities is available in both a CPU and GPU form. To change between the two the user needs only change a keyword argument in the \texttt{Python}-based user interface.

\subsection{Fast Trajectory Module}\label{sec:traj}

The initial part of EMRI waveform creation is the determination of the CO's phase space trajectory. This orbit-averaged trajectory is determined over time in terms of $G(t):=\{p(t), e(t), x_I(t), \Phi_\varphi(t), \Phi_\theta(t), \Phi_r(t)\}$. These quantities satisfy (coupled) ordinary differential equations (ODEs) that we solve via numerical integration. As will be discussed in Sections \ref{sec:schwarz_ecc} and \ref{sec:newAAK}, the ODEs are specific to each trajectory implemented. For a flux-driven trajectory, these time derivatives are determined from $\{\dot{E}, \dot{L}, \dot{Q}\}$, where these three quantities represent, in order, the time derivatives of the orbital energy, axial orbital angular momentum, and Carter constant. These quantities can be determined from the Teukolsky amplitudes $Z^{\rm H,\infty}_{lmkn}$ \cite{Hughes2000, Hughes2021EMRI_FD}. In the present work, we implement a flux-driven trajectory while post-adiabatic trajectories are developed.

The computation of the amplitudes $Z^{\rm H,\infty}_{lmkn}$ is too slow to be implemented on the fly and, therefore, impractical for any trajectories generated for data analysis. These quantities will instead need to be pre-computed on a grid in $\{a, p, e, x_I\}$ space and matched with a fast interpolation method for online calculations. In \mbox{Section \ref{sec:schwarz_ecc}}, we will discuss the specifics of the first implementation of this scheme in the Schwarzschild eccentric regime. 

For the numerical integration, we employ an adaptive eighth-order Runge-Kutta integrator from the GNU Scientific Library (GSL) \cite{GSL}. The integrator outputs the trajectory at a small number of time steps ($\sim100$) as the trajectory is very smooth. This is a key component of the overall waveform speed and architecture. These sparsely sampled trajectories are heavily leveraged so as to reduce intermediate computations prior to the final waveform build. These sparse trajectories, when upsampled with a cubic spline, show negligible absolute error when compared to densely sampled trajectories with small fixed timesteps. These methods will also be applicable to post-adiabatic trajectories, but adjustments will be required to incorporate resonances.

For maximal flexibility and verification, the base trajectory module in the FEW package allows the user to customize the trajectory output. For example, users can densely sample their trajectories; resample the trajectories using cubic splines; and/or  convert from coordinate to dimensionless time scaled by the mass of the MBH. Due to the serial nature of trajectory computations, these modules are currently only implemented on the CPU. 

\subsection{Amplitude Module}\label{sec:amp}

As stand-alone parts, amplitude modules take in $\{a, p, e, x_I\}$ in vectorized form, returning a two-dimensional array of individual complex mode amplitudes, $A_{lmkn}$, for each parameter set provided. In a Schwarzschild eccentric background, the harmonic basis consists of 3843 harmonic modes. When expanding to the generic Kerr regime, the polar harmonic index $k$ will be introduced, and the number of modes will expand by roughly an order of magnitude. In \mbox{Section \ref{sec:analysis}}, we show that waveforms with purposefully and strategically reduced harmonic content, in many situations, will be sufficient for data analysis purposes.


Within a larger waveform model, the amplitude module will take a specific set of orbital parameters given by the trajectory outputs. These will generally be sparse trajectories that will produce sparse amplitude computations at the same time cadence as the input trajectories. Since amplitudes vary slowly on the radiation reaction timescale their spare sampling is once again an important component of maintaining the speed of the overall waveforms (see \mbox{Figure \ref{fig:timing}}). 

The amplitudes $A_{lmkn}$ are determined for each orbital parameter set, independent of all other orbit points along a trajectory.  This facilitates parallelizable computations for the amplitude modules; where methods are parallelizable, they will be available in both a GPU and CPU option.

\subsection{Waveform Summation Module}\label{sec:summation}

With the constituent parts in place, the waveform summation module takes trajectory and amplitude information and forms the final waveform. This is done by upsampling the sparse arrays to the true sampling of the data stream. This is a crucial step in the waveform creation process; even with various improvements described in this paper, it remains the bottleneck in terms of accelerated EMRI computations. For this reason, as well as the parallelizable nature of this large-dimensional computation, accelerator hardware is necessary to bring the compute time of this process down to reasonable levels for data analysis. 

There are two types of summation modules provided in the FEW package. They are both generic to most waveforms that will be implemented in the FEW model (an exception to this is the new AAK discussed in Section~\ref{sec:newAAK}). The more basic option provided performs a direct summation with amplitude and trajectory information at the array density provided. This type of summation is slower than the second type of summation, and is usually employed when densely stepping trajectories and amplitude calculations. 

The faster waveform summation is performed by using cubic splines to interpolate the sparse array information. Therefore, this is referred to as the "Interpolated Summation." This process begins with a special cubic spline interpolation implemented for efficiency on GPUs. It leverages \texttt{cuSparse} from the NVIDIA library to quickly compute spline coefficients in a tridiagonal banded matrix representation. The specific spline implemented is the ``not-a-knot'' spline in terms of its boundary conditions. This spline implementation assumes all arrays to be interpolated exist on the same sparse time grid. The parallelized spline fitting allows us to quickly determine the spline coefficients simultaneously for all phases and individual mode amplitudes ($\sim10^2-10^4$ separate splines based on overall mode content). This cubic spline, available in both CPU and GPU versions, is also a stand alone piece of the FEW framework and can be leveraged in other projects where similar spline computations are needed. 

Once the spline coefficients are determined, the actual summation kernel evaluates the spline at the sampling rate of the output data stream. Then, the output phase and amplitude values are combined within each mode according to Eq.~\eqref{eq:main_wave}. In this step, we exploit a symmetry: the amplitude relation
\begin{equation}\label{eq:amp_symmetry}
    A_{l,-m,-k,-n} = (-1)^{(l+k)}A^*_{lmkn}\;,
\end{equation}
where the superscript $^*$ denotes complex conjugation.  This allows us to only compute modes for $m\geq0$; modes with $m < 0$ can inferred using Eq.\ (\ref{eq:amp_symmetry}).  Once the final complex contribution to the waveform is determined at each specific time value within each mode, the modes in a single time step are combined. This is effectively the cause of the high-dimensional computational difficulty of the problem. For a year long data stream sampled at 0.1 Hz, a typical sampling rate chosen for LISA data, this creates a summation across $\sim10^2-10^4$ modes at each of the $\sim3.15\times10^6$ time steps. As can be seen in \mbox{Figure \ref{fig:timing}}, the GPU accelerator makes this step tractable for data analysis with LISA. 

\subsection{Utility Modules}\label{sec:utility}

In addition to the base modules discussed above, the FEW package provides many utility modules that are important to the overall waveform creation model, including, but not limited to, a separatrix calculator in generic Kerr spacetime; a generic Kerr fundamental frequency calculator; spin-weighted spherical harmonics; and a basic mismatch calculator. The interested reader can find documentation on these utilities in the FEW package \cite{BHPToolkit-FEW}.

The most important of these modules that is not specific to any individual waveform model is the ``Mode Selector'' module. The Mode Selector performs an online calculation to determine which of the individual harmonic modes contribute power to within a user-defined threshold of the total power emitted by all modes (denoted as $\epsilon$). This begins by taking arrays of the power within each mode at each sparse time step determined from the complex amplitudes output by an amplitude module. It then operates entirely within each time step. It sorts the individual modes in descending order and computes a cumulative summation. From this sorted array, it is determined where the additional power contributed by an individual mode falls below a user provided threshold  related to the total power emitted ($\epsilon$) at that time step (the final entry in the cumulative sum). This gives a set of contributing modes within each time step. We then take the union of all contributing modes across all time steps to maintain continuity across time. There is a very small loss of accuracy in this procedure since we are not performing this calculation at each and every dense time step. However, this would only lead to the inclusion of fringe modes that contribute only slightly to the overall waveform power. As indicated by the overall waveform mismatch determined empirically in \mbox{Section \ref{sec:mismatch}}, this loss in accuracy is negligible across our parameter space. Within the Mode Selector module is the option to include noise-weighting. We find this does not make a major difference in terms of the specific modes chosen and do not use this method in this paper. 

\subsection{Overall Code Design}

The FEW code was carefully designed to enhance its modularity, flexibility, ease-of-use, and acceleration capabilities. Here, we will discuss the overall code structure from the top user interface down. 

The user accesses everything in FEW through a \texttt{Python} interface. At the highest level is a generic waveform generator where the specific waveform is provided as an argument during instantiation. This generic interface allows for seamless transitions between waveform backends for fast waveform comparisons. 

Below the top-level generic generator are the specific waveform models. We currently provide three complete waveform models: \texttt{FastSchwarzschildEccentricFlux} (\mbox{Section \ref{sec:schwarz_ecc}}), \texttt{SlowSchwarzschildEccentricFlux} (\mbox{Section \ref{sec:slow}}) and \texttt{Pn5AAKWaveform} (\mbox{Section \ref{sec:newAAK}}). The first two are provided in the source frame. The latter is built in the detector frame. The generic interface accounts for this difference when comparing two different generation frames (see \mbox{Section \ref{sec:detector_frame}}). 

Each of these waveforms is then broken down into \texttt{Python} modules as detailed above. Within the full waveforms and their constituent modules, \texttt{Python} is leveraged for providing the overall code and package structure, performing simple operations, and preparing input for the low-level \texttt{C++}/\texttt{CUDA} code. Each module containing a calculation in \texttt{C++}/\texttt{CUDA} uses a thin, customized \texttt{Cython} interface \cite{Cython, CUDAwrapper}. 

At the lowest level lives the \texttt{C++}/\texttt{CUDA} functions that are built for speed. Our main goal is to limit the amount of code in \texttt{C++}/\texttt{CUDA} to only necessary bits that require speed and enhanced acceleration capabilities. The implementation of parallelization and acceleration capabilities is provided in both \texttt{Python} modules and \texttt{C++}/\texttt{CUDA} code. In \texttt{Python}, CPU/GPU agnostic code is designed using \texttt{NumPy} \cite{Numpy} and \texttt{CuPy} \cite{CuPy}. In \texttt{C++}/\texttt{CUDA}, functions are first designed for GPUs to ensure enhanced performance and acceleration and are then adapted for CPUs using \texttt{OpenMP} \cite{dagum1998openmp}. The code is specifically designed to minimize duplication of code across the different hardware, leveraging compiler directives specific to each device only when necessary.

\section{Adiabatic Schwarzschild Eccentric Waveforms}\label{sec:schwarz_ecc}

For the development of the first fully relativistic EMRI template waveforms, we focused on the initial task of performing these computations for eccentric inspirals into a Schwarzschild black hole. In this regime, we can exclude various parameters from the full EMRI description. The MBH is non-rotating allowing us to set $a=0$ and the spacetime is spherically symmetric. This leads to the removal of the inclination parameter ($x_I=1$ for completeness) as we can consider any orbit to be in the equatorial plane. This also allows for the removal of polar phases ($\Phi_\theta$) and indices ($k$) from Eq.~\eqref{eq:main_wave} reducing the summation to a sum over the $lmn$ indicies. Combining the spheroidal harmonic $S_{lmkn}(t, \theta)$ with $e^{im\phi}$ and then taking the limit in the Schwarzschild background reduces these terms to the regular -2 spin-weighted spherical harmonics, $_{-2}Y_{lm}(\theta, \phi)$ \cite{Teukolsky1973}. Eq.~\eqref{eq:main_wave} in the Schwarzschild eccentric regime then becomes:
\begin{equation}
    h = \frac{\mu}{d_L}\sum_{lmn} A_{lmn}(t)_{-2}Y_{lm}(\theta, \phi)e^{-i\Phi_{mn}(t)}.
\end{equation}
Notice this means the angular harmonic term is no longer time dependent, but constant over the whole orbit (see \mbox{Section \ref{sec:discuss}} for more discussion). 

Here we will detail how specific implementations of the various modules presented in \mbox{Section \ref{sec:FEW_framework}} were combined to create the first fast and fully relativistic EMRI waveforms. 
With the trajectories and amplitudes determined according to the following sections, this information was combined with the spin-weighted spherical harmonics and then put through mode selection, both of which are described in \mbox{Section \ref{sec:utility}}. Following mode selection, the waveform is built using the interpolated summation (see  \mbox{Section \ref{sec:summation}}). This eccentric Schwarzschild adiabatic model is valid for $p_\text{min} \leq p \leq p_s + 10$ and $0 \leq e \leq 0.7$, where $p_s$ is the separatrix \cite{Stein:2019buj} and $p_\text{min} = \text{max} (p_s + 0.1, 7p_s - 41.9).$

\subsection{Flux-driven Trajectories}\label{sec:flux-driven}

The first specific piece of a relativistic waveform is a relativistic trajectory. This module was built under the generic fast trajectory formula described in \mbox{Section \ref{sec:traj}}. For this waveform, we operate with a flux-based adiabatic trajectory. To do this we must calculate $(\dot{p},\dot{e})$ from $(\dot{E}, \dot{L})$.  The energy and angular momentum flux must be calculated quickly for generic eccentric geodesics (limited to our domain of validity) in a Schwarzschild background as the numerical integrator evolves the trajectory forward in time. This is not possible to do by directly computing Teukolsky amplitudes and fluxes because the duration of this calculation is orders of magnitude larger than the time we require for the entire trajectory ($\sim$ms). We therefore rely on accurate and efficient interpolation techniques. 

The first step to evaluating the trajectories is to compute a grid of flux values using a Teukolsky code. In order to efficiently interpolate the fluxes with bicubic splines, it is helpful to place place the data on a grid with uniform spacing. As in Schwarzschild spacetime the separatrix is given by  $p_s=6 + 2e$ \cite{Stein:2019buj}, we instead introduce $u=\ln(p-p_s+3.9)$ \cite{Chua2020RapidGenLetter}. This allows us to build a uniform grid in $(u, e)$ space with $1.37\leq u \leq 3.82$ in steps of 0.05 and $0.0\leq e \leq 0.8$ in steps of 0.025. This parameterization allows for more points closer to the separatrix where the orbital quantities vary more rapidly. The grid coordinate $u$ corresponds to separations of $p_s + 0.03\leq p \leq p_s+41.6$. 

Rather than a two-dimensional interpolation over this grid of the actual flux values, we subtract out the leading PN behavior and instead interpolate over an effective flux residual to reduce error in the interpolation. We then construct bicubic splines over $(u,e)$ of $(\dot{E}_\text{spl}, \dot{L}_\text{spl})$ given by 
\begin{align}
    \begin{split}
    \dot{E}_\text{spl} &= (\dot{E} - \dot{E}_\text{PN})\Omega_\varphi^{-4} \quad \text{and} \\
    \dot{L}_\text{spl} &= (\dot{L} - \dot{L}_\text{PN})\Omega_\varphi^{-3},
    \end{split}
\end{align}
where the PN behavior is given by \cite{Munna:2020juq} 
\begin{align}
    \begin{split}
    \dot{E}_\text{PN} &= \frac{(96 + 292e^2 + 37e^4)}{15(1 - e^2)^{7/2}}\Omega_\varphi^{10/3} \quad \text{and} \\
    \dot{L}_\text{PN} &= \frac{4(8 + 7e^2)}{5(e^2 - 1)^2}\Omega_\varphi^{7/3}.
    \end{split}
\end{align}
At evaluation time, the integrator determines the fundamental frequencies ($\Omega_\varphi, \Omega_r$) \cite{Cutler1994RadiationSchwarzschild} based on $(p,e)$, followed by the PN contribution to ($\dot{E},\dot{L}$). It then converts ($p,e$) to $(u,e)$ and evaluates the interpolant to get the effective adiabatic residual of ($\dot{E},\dot{L}$). With ($\dot{E}, \dot{L}$) in hand, ($\dot{p},\dot{e}$) are computed using \cite{Cutler1994RadiationSchwarzschild}. 

The integrator begins at $(p_0, e_0, \Phi_{\varphi,0}, \Phi_{r,0})$ and integrates until it takes a step that is within 0.1 of the separatrix. As the integrator steps over $p_s + 0.1$, the integrator reverts back to its previous value and walks in smaller steps until it reaches within $10^{-8}$ of $p_s + 0.1$. Therefore, the trajectories that reach the separatrix end at $p\approx p_s + 0.1 + 10^{-8}$. Without this small stepping operation at the end, trajectories finish at whatever point the integrator finds within $p_s + 0.1$. This causes inconsistencies in the time and separation at the end of the waveform. Before considering this effect, we found the likelihood computations in \mbox{Section \ref{sec:posteriors}} to be noisy and not smoothly varying as would be expected over small scales in parameter space. 

As mentioned in \mbox{Section \ref{sec:traj}}, this integration produces $\sim100$ points along the trajectory as $(p(t), e(t), \Phi_\varphi(t), \Phi_r(t))$. An example of the trajectories in $(p,e)$ space can be seen in \mbox{Figure \ref{fig:mismatch}}.

\subsection{RomanNet Amplitude Generation}\label{sec:RomanNet}

Following the generation of orbital parameter trajectories, the complex amplitudes of the many harmonic modes must be produced. To accomplish this task, we implement a version of a reduced-order-model with artificial neurons (ROMAN) \cite{Chua:2018woh} within the amplitude module framework discussed in \mbox{Section \ref{sec:amp}}.

Complex amplitudes were generated along the same grid as the total flux values discussed in \mbox{Section \ref{sec:flux-driven}}. In our mode set, we analyze $l\in[2,10]$, $m\in[0, l]$, and $n\in[-30, 30]$, which totals 3843 modes. Mode amplitudes with $m<0$ were determined using Eq.~\eqref{eq:amp_symmetry} with $k=0$. To prepare this group of 3843 modes per $(u,e)$ pair for use in our neural network amplitude generation scheme, we first compress the information using the method of reduced order modeling (ROM) \cite{Field:2011mf}. ROM is a powerful technique that is used within gravitational wave analysis to build efficient surrogate waveforms \cite{Field:2013cfa}, and to accelerate likelihood calculations through the method known as reduced-order-quadrature (ROQ) \cite{Canizares:2013ywa}. ROMAN as originally proposed fulfils the same function as surrogates + ROQ, by using neural networks as waveform fits and working natively in the reduced-order domain. Here, however, we use ROMAN to fit the set of mode amplitudes instead, which leverages the strength of neural networks for global regression over high-dimensional complex spaces.

With the 3843 modes per $(u,e)$ cast as a vector $A_i=\text{vec}(A_{lmn})\in\mathbb{C}^{3843}\simeq\mathbb{R}^{7686}$, we compress our dataset using a greedy algorithm from the \texttt{Python} package \texttt{RomPy} \cite{RomPy}. The data is compressed to a reduced basis $B$ representing the span of the uniform grid of complex amplitude data. The reduced-order basis is represented by reduced-order basis coefficients, $\alpha_j$:
\begin{equation}
    A_i(u, e) = \sum_j \alpha_j(u,e)B_{ji}\equiv\alpha_j(u,e).
\end{equation}
The data is compressed by a factor of $\sim40$ with $\alpha_j\in\mathbb{C}^{99}\simeq\mathbb{R}^{198}$.

We then trained the neural network with inputs equivalent to the grid of $(u,e)$ values and outputs equivalent to $\alpha_j\in\mathbb{R}^{198}$. The neural network architecture is a basic fully-connected neural network. It consists of 20 hidden layers. The first hidden layer has 4 nodes with the subsequent five layers increasing each node count by a factor of 2 until the node count reaches 256. The remainder of the hidden layers all employ 256 nodes. To incorporate non-linear behavior we activate each hidden layer with a Leaky ReLU function \cite{Maas2013RectifierNI} (the output layer is not activated). The training is performed in minibatches \cite{Goodfellow-et-al-2016} of 810 with the ADAM gradient descent optimizer \cite{ADAM}. The loss function is the standard $L^2$ loss function: $L = \langle|\alpha - \hat{\alpha}|^2\rangle$, where $\langle\cdot\rangle$ represents the average over a minibatch and $|\cdot|$ is Hermitian. The neural network is trained over $3\times10^4$ epochs. 

At run-time, the RomanNet module is given the arrays of $(p,e)$ determined from the trajectory module. It then converts these pairs to $(u, e)$ and inputs these values into the trained neural network. The network outputs $\alpha_j$, which are then transformed back to the full amplitude space with $\alpha\cdot B$. The amplitude vectors output by the module are then renormalized to a more accurate vector norm generated by a bicubic spline during and output by the trajectory module at each time step.

\subsection{Slow Reference Waveform}\label{sec:slow}

To verify that our new waveforms are accurate, we must compare against more accurate, slowly generated waveforms. Generally, waveforms produced directly from the modelling community are too slow to produce $\sim$year of consecutive data, which is the duration needed for proper data analysis-related tests. Therefore, we have constructed a ``slow'' version of our Schwarzschild adiabatic waveform by implementing trajectory, amplitude, and summation modules that focus on accuracy rather than speed. 

The trajectory for this comparison waveform is determined in the same manner as the fast trajectory modules in terms of evolving the trajectory forward in time from $(p_0, e_0)$ to within $\sim0.1$ of the separatrix. The difference is that the time steps are fixed to the time step in the data stream ($\sim$10s). This allows for a maximally accurate calculation of the orbital and phase trajectories without performing any large steps via adaptive methods. 

The amplitudes are determined by using the same set of $A_{lmn}(u,e)$ used for the underlying training set of the neural network. However, rather than using advanced methods for fitting the data, we instead settle for a simple, but highly accurate bicubic spline over the real and imaginary pieces of each individual mode. This is inefficient in terms of memory storage and speed of evaluation, but is necessary to create highly accurate generic EMRI waveforms to compare against our more advanced generation methods. Additionally, amplitudes are not calculated at sparse points, but rather at each point in the dense trajectory output by the slowly evolving trajectory module just previously mentioned. 

The final waveform summation is calculated using the direct summation method described briefly in \mbox{Section \ref{sec:summation}}. This is performed by combining the phase and amplitude information at each time point in the data stream without interpolating any of these quantities.

\section{An Improved Augmented Analytic Kludge}\label{sec:newAAK}

Another addition to the FEW framework is a new version of the Augmented Analytic Kludge (AAK) first presented in \cite{Chua2015, Chua2017}. AAK waveforms remain useful even with fast relativistic waveforms under development, since they are extensive in their coverage of the generic Kerr parameter space by construction. These waveforms are still useful in ongoing data analysis studies for LISA to understand the extraction of the spin of the MBH, as well as the complexity associated with generic orbit configurations. Here, rather than detailing the AAK formalism, we will describe what has changed in our new version of the AAK. We refer the interested reader to \cite{Chua2015, Chua2017} to understand the foundations for generating AAK waveforms. 

The new AAK effectively glues together a more accurate and robust trajectory module with the old AAK's waveform generator (or summation module in the FEW framework). The new trajectory, built in the fashion of the fast trajectory module described in \mbox{Section \ref{sec:traj}}, integrates through the parameter evolution using 5PN flux values for derivatives $\{\dot{p}, \dot{e}, \dot{Y}\}$ ($Y=\cos\iota$) \cite{Fujita:2020zxe}. Note that $Y=\cos{\iota}\equiv L_z/\sqrt{L_z^2 + Q}$ where $\iota$ is different from the angle $I$ used in this paper to describe the orbital inclination angle in the relativistic construction. The parameter $\iota$ is included here because it is explicitly used in the semi-relativistic formulation. For the time derivatives of the phases, we employ the same fundamental frequencies as in the relativistic waveforms, $(\Omega_\varphi, \Omega_\theta, \Omega_r)$, which now includes the polar frequency, $\Omega_\theta$, required in the generic Kerr regime. This indicates an important update to the original AAK waveform: $\iota$ is now an evolving quantity whereas in prior versions it was taken to be a constant. Additionally, a key distinction here with respect to the Schwarzschild trajectory described in \mbox{Section \ref{sec:flux-driven}} is that the separatrix is no longer simply given by $p_s=6 + 2e$. Instread, we employ a numerical computation of the separatrix that depends on $(a,e,Y)$ given in \cite{Stein:2019buj}. We implement this calculation in \texttt{C++} leveraging the \texttt{GSL} library \cite{GSL} for root finding. Once the trajectory has reached within 0.1 of the separatrix, it stops and performs the finishing integration step described for the relativistic trajectory in \mbox{Section \ref{sec:flux-driven}}. This higher-dimensional trajectory computation still outputs arrays that are $\sim100$ points in length. 

The old AAK built a trajectory by using frequency evolution from the Numerical Kludge \cite{Babak2007} and mapping it onto the frequency basis used in the original Analytic Kludge \cite{BarackCutler2004}. This actually resulted in a time evolving effective mass and spin of the MBH, along with a few adhoc additions within the waveform building step that ensured the AAK would maintain roughly the same frequency evolution as the NK. 

With the new 5PN trajectory, we do away with the mapping and adhoc steps within the waveform build. Instead, we directly calculate the fundamental frequencies along the trajectory and convert these to the basis used in the AAK waveform generation step:
\begin{align}
    \begin{split}
    \dot{\Phi}&= \Omega_r, \\
    \dot{\gamma} &= \Omega_\theta - \Omega_r, \quad \text{and}  \\
    \dot{\alpha} &= \Omega_\varphi - \Omega_\theta,
    \end{split}
\end{align}
where $\dot{\Phi}$ is the rate of change for the quasi-Keplerian mean anomaly, $\dot{\alpha}$ is the rate of Lense-Thirring precession, and $\dot{\gamma}+\dot{\alpha}$ is the angular rate of periapsis precession. The phases in this basis, $\{\Phi=\Phi_r, \gamma, \alpha\}$, are given by $\int_0^t dt' (\dot{\Phi}(t'),\dot{\gamma}(t'),\dot{\alpha}(t'))$. In the original AAK, a calculation was performed at each time step to determine the proper $\Omega_\varphi$. Since we already have this quantity, we have removed this computation and instead feed the $\Omega_\varphi$ array directly into the waveform summation step.

Similar to the interpolated summation described in \mbox{Section \ref{sec:summation}}, all time evolving information necessary for the final waveform summation is interpolated with a cubic spline. After preparing all the necessary quantities along a sparse trajectory, the quantities that require interpolation are $(e, \iota, \Phi, \gamma, \alpha, \dot{\Phi}, \dot{\gamma}, \Omega_\varphi)$ ($p(t)$ information is included via $\dot{\Phi}, \dot{\gamma}$, and $\Omega_\varphi$). The actual waveform summation step is exactly the same as the old AAK, with the exception of the interpolation of these quantities and the aforementioned direct input of $\Omega_\varphi$.

The new AAK summation is coded as a separate module from the actual trajectory determination. Therefore, in sticking to the FEW framework, these modules are interchangeable and usable as stand-alone tools. If a user builds their own trajectory in generic Kerr, they can attach this AAK summation to it in the current absence of a fully generic relativistic Kerr model. This waveform summation module is also GPU-acccelerated in keeping with the CPU/GPU agnostic nature of the FEW framework. 

\section{Gravitational Wave Analysis}\label{sec:analysis}

A key difficulty with the lack of fast EMRI waveforms was the inability to perform detailed data analysis studies in a tractable amount of time. We show how the accelerated FEW framework can be leveraged to study EMRI systems including basic underlying EMRI waveform information, as well as tests of EMRI detectability and parameter characterization. A variety of fast FEW waveforms were studied with differing settings: one waveform uses a mode content parameter set to $10^{-5}$ (FF$\epsilon$5) to represent a relativistic waveform with a large amount of harmonic modes; another with relativistic mode content, but a small number of modes with $\epsilon$ set to $10^{-2}$ (FF$\epsilon$2); and a relativistic quadrupolar waveform with only the $(l,m,n)=(2,2,n)$ modes (FF22). Additionally, we tested a waveform that combines the trajectory module from the fast relativistic Schwarzschild waveform with the AAK waveform summation module. This produces a waveform with a relativistic adiabatic trajectory and AAK-generated amplitudes limited to the Schwarzschild eccentric regime. This means when we compare this Schwarzschild AAK (SchAAK) waveform with the fast Schwarzschild FEW models, the phase trajectories are \textit{exactly} the same. Therefore, the focus of these comparisons is on the amplitude difference between the models in order to understand how accurate the amplitudes of EMRIs need to be in practice to perform data analysis. Future studies with these models will help illuminate these questions. Here, we provide an initial look at the Schwarzschild eccentric adiabatic regime.

We use the generic FEW waveform interface to produce detector-frame waveforms with the same conventions across all models. While these waveforms are produced in the detector frame, we do not use a LISA response function as accurate versions are not yet available; therefore, we focus on intrinsic parameters during any comparisons. All angular quantities are chosen one time and used for all tests. They are chosen from a random uniform distribution across their domain. The values tested were $(\theta_S, \phi_S, \theta_K, \phi_K, \Phi_{\varphi,0}, \Phi_{r, 0}) = (0.54, 5.36, 1.73, 3.20, 3.23, 4.72)$. As we are in the Schwarzschild eccentric regime, we set $(a, Y_0, \Phi_{\theta, 0})=(0.0, 1.0, 0.0)$. All waveforms are two years in length with a measurement cadence of 10 seconds. 

The following studies employ common gravitational-wave metrics to further understand the use of relativistic waveforms in EMRI analysis. The conventional gravitational-wave likelihood $\mathcal{L}$ between a set of strain data, $d(t)$, and a template waveform, $h(t)$, is given by
\begin{equation}
    \ln{\mathcal{L}} = -\frac{1}{2}\langle d - h| d - h\rangle\;,
\end{equation}
where we have introduced the noise-weighted inner product:
\begin{equation}
    \langle a|b \rangle =  4\text{Re}\left[\int_{0}^{\infty} \frac{\tilde{a}(f)*\tilde{b}(f)}{S_n(f)}df\right]\;.
\end{equation}
Here, $\tilde{a}(f)$ is the Fourier transform of $a(t)$, and $S_n(f)$ is the power spectral density (PSD) of the noise. The SNR of a given source is equivalent to $\sqrt{\langle d | h\rangle}$. The analysis in \mbox{Section \ref{sec:posteriors}} is based on the LISA mission. Therefore, the PSD used was the ``SciRDv1'' version of the LISA noise curve \cite{SciRD1} without Galactic binary foreground noise for convenience and to focus on the waveform-model bias. \mbox{Sections \ref{sec:harmonic_analysis}} and \ref{sec:mismatch} instead focus directly on the waveform model, setting the PSD to 1 for all frequencies to avoid noise-weighting for a specific detector. For these computations we use the ``overlap'' or ``mismatch'' $= 1 - $ overlap. The overlap is a normalized inner product:
\begin{equation}
    \text{overlap}(a,b) = \frac{\quad\langle a|b \rangle}{\sqrt{\langle a|a \rangle \langle b|b \rangle}}.
\end{equation}

\subsection{Harmonic Mode Analysis}\label{sec:harmonic_analysis}

\begin{figure*}
\begin{center}
\includegraphics[scale=0.7]{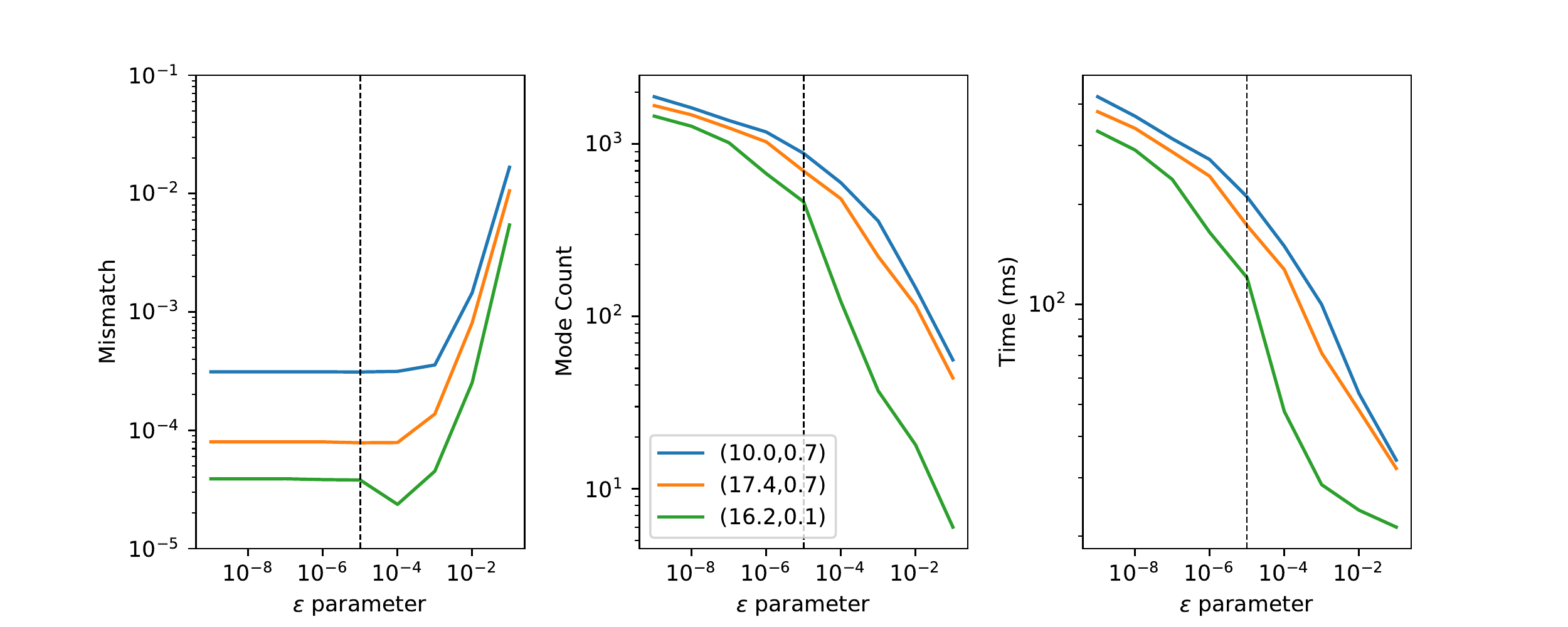}
\caption{The effect of adjusting the $\epsilon$ parameter on the mismatch, mode count, and waveform generation time. This parameter determines the threshold for the fractional added power of the harmonic modes. For each plot shown, three sets of parameters are run, labelled in the legend of the central plot as $(p_0,e_0)$. The left graph shows the mismatch versus $\epsilon$ compared against slow FEW waveforms (\mbox{Section \ref{sec:slow}}). The mismatch asymptotically approaches a minimum value specific to each set of parameters. This minimum value above zero is due to the noisy floor of the RomanNet. The center and right plots show the corresponding number of modes and speed, respectively, for each set of parameters tested. A dashed vertical line at $\epsilon=10^{-5}$ is added to each plot to indicate the $\epsilon$ chosen for the base injection waveform in the following sections.
}\label{fig:mode_content}
\end{center}
\end{figure*}

The overall accuracy and speed of the FEW waveform is user controlled by setting the fractional mode power parameter, $\epsilon$, as described in Section \ref{sec:utility} above. This controls the number of modes built into the waveform. Mismatch values, mode count, and generation speed across $\epsilon$ values are shown in \mbox{Figure \ref{fig:mode_content}}. Three points representing the boundary of the FEW domain of validity are shown: $(p_0, e_0)\in[(10, 0.7), (17.4, 0.7), (16.2, 0.1)]$. The mismatch behavior across waveforms is effectively the same. At the high end near $\epsilon=10^{-1}$, the mismatch is $\sim10^{-2}$. The mismatch decreases in a power-law behavior as $\epsilon$ is decreased until $\epsilon\sim10^{-5}$. It then asymptotically approaches $\sim10^{-4}$ as $\epsilon$ tends toward zero. This behavior is due to the effective noise floor of the ROMAN network amplitudes at about a factor of $\sim10^{-6}$ relative to the maximum amplitude mode. The slight dip in mismatch for $(p_0,e_0)=(16.2, 0.1)$ at $10^{-4}$ is also due to this noise in the neural network amplitudes. For these lower eccentricity sources, where power is concentrated in fewer modes, smaller $\epsilon$ values result in many noisy (albeit low power) modes added to the waveform. At $\epsilon=10^{-4}$, the noisy mode contribution is minimized while the true mode contribution is maximized.

The mode count behavior is similar across initial parameters. The lowest eccentricity source shown with $(p_0,e_0)=(16.2, 0.1)$ has 1453 modes in the waveform at $\epsilon=10^{-9}$. At $\epsilon=10^{-1}$ it contains only 6 modes. The system with $(p_0,e_0)=(10, 0.7)$ contains 1883 and 56 modes with $\epsilon=10^{-9}$ and $\epsilon=10^{-1}$, respectively. The waveform generation speed follows closely with the mode content because this mode count in the waveform summation is the most important contribution to the overall timing. This is discussed further in \mbox{Section \ref{sec:timing}}.

\begin{figure*}
\begin{center}
\includegraphics[width=\textwidth]{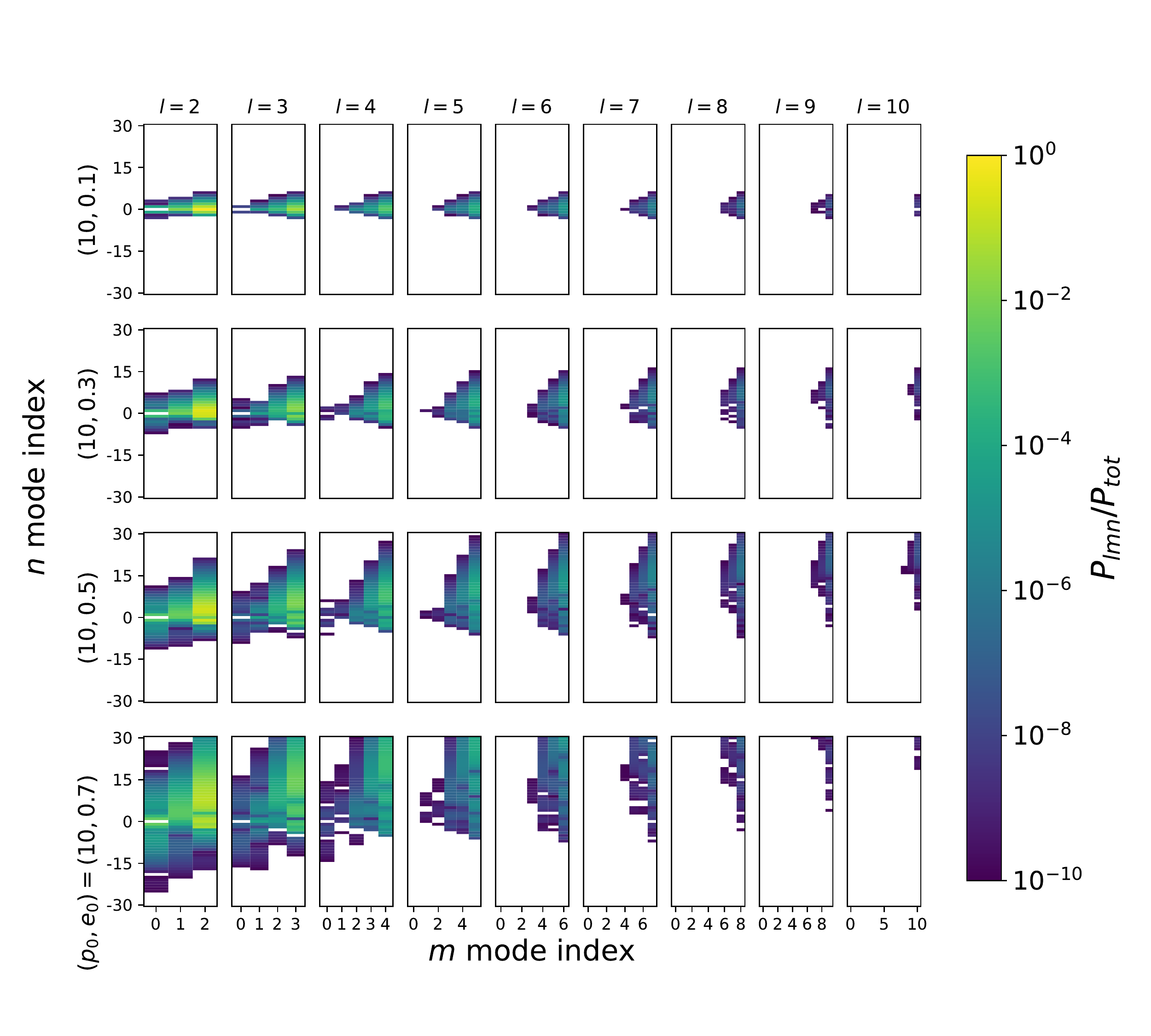}
\caption{The mode power content as a fraction of the total power emitted for a given geodesic orbit with various eccentricities and $p_0=10$. The mode data was computed with bicubic spline-generated amplitudes (see \mbox{Section \ref{sec:slow}}). Plots in each row have the same eccentricity (labelled along the left edge). Columns represent a singular $l$ mode value (labelled along the top edge). Within each plot, $m$ and $n$ mode values are given along the horizontal and vertical axes, respectively. We show $n\in[-30,30]$ and $0 \leq m \leq l$. For $m<0$, the plot would be the mirror image of above reflected around $n=0$. Modes with $P_{lmn}/P_\text{tot}<10^{-10}$ are not shown.
}\label{fig:visual}
\end{center}
\end{figure*}

The harmonic structure of relativistic EMRI waveforms is rich and complex. Even in Schwarzschild eccentric where the $k$-indexed polar modes are ignored, there is not a clear set of relations to succinctly describe the structure. This may be a topic of future work. However, the mode structure can be qualitatively visualized. \mbox{Figure \ref{fig:visual}} shows a gridded visualization of the power in every $(l,m,n)$ mode for different values of eccentricity (the $\epsilon$ mode selection parameter was not employed here). The grid structure is explained in the caption. Note that $m<0$ modes are not included because the power is equivalent in the $\pm m$ modes (if angular harmonics are not included). The power in each mode is shown as a fraction of the total power until the fractional mode power falls below $10^{-10}$. This visualization is created with the bicubic spline amplitude calculator to make sure to remove any noisy, lower power modes. As the eccentricity is increased, the mode power spreads out to a larger number of modes as well as to higher $n$ modes (or lower $n$ modes for $m<0$). The contributing $n$ modes within each $l$ subset also tend towards a smaller number of $n$ modes located at larger $n$ values as $l$ is increased.

\begin{figure*}
\begin{center}
\includegraphics[width=\textwidth]{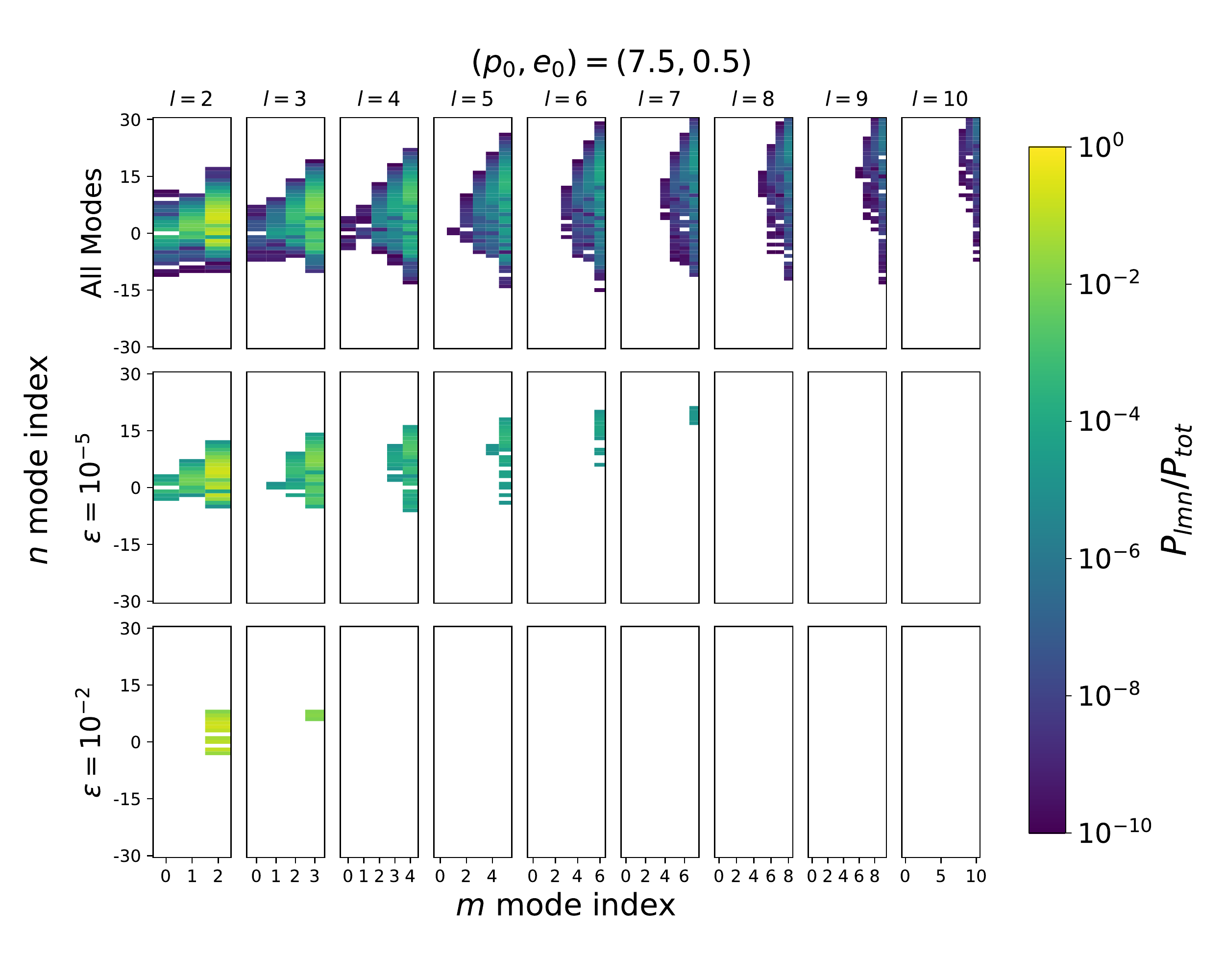}
\caption{Similar to \mbox{Figure \ref{fig:visual}} but for $(p_0,e_0)=(7.5, 0.5)$ and different values of $\epsilon$ which controls the mode content in the waveform. Geodesic output from all modes, modes selected with $\epsilon=10^{-5}$, and modes selected with $\epsilon=10^{-2}$ are shown in each row from top to bottom, respectively. In \mbox{Section \ref{sec:posteriors}}, we show waveforms built with $\epsilon=10^{-2}$ show high fidelity against more accurate waveforms indicating building waveforms with less modes will increase speed without sacrificing much accuracy.  }\label{fig:visual2}
\end{center}
\end{figure*}

An additional mode content visualization is shown in \mbox{Figure \ref{fig:visual2}}. However, this visualization fixes $(p_0,e_0)=(7.5, 0.5)$. Instead of varying parameters, it shows how the mode selection parameter $\epsilon$ affects the mode content from a singular geodesic. When building a waveform, mode selection is performed for each instantaneous geodesic in the sparse trajectory. However, please note the final waveform is constructed with the union of all selected modes across the entire sparse trajectory. The top row in the figure shows the fractional mode content in all modes. Many ($l,m,n$) modes are highlighted in this highly relativistic orbit. When selecting modes with $\epsilon=10^{-5}$, all modes with $l>7$ are eliminated from consideration with the strongest modes in $l\leq7$ remaining. For $\epsilon=10^{-2}$, a much higher fraction of modes are eliminated with only $l\leq3$ modes remaining. However, these $l=3$ modes are crucial for comparison against quadrupolar ($l=m=2$) waveforms like the AAK (see \mbox{Sections \ref{sec:mismatch}} and \ref{sec:posteriors}).

\subsection{Mismatch Analysis}\label{sec:mismatch}

In the original FEW model paper \cite{Chua2020RapidGenLetter}, we performed an initial analysis on the mismatch of the FEW model over time. In this section, we expand on that analysis to test more waveforms against the slow FEW Schwarzschild eccentric waveform presented in \mbox{Section \ref{sec:slow}}. 

\begin{figure*}
\begin{center}
\includegraphics[scale=0.45]{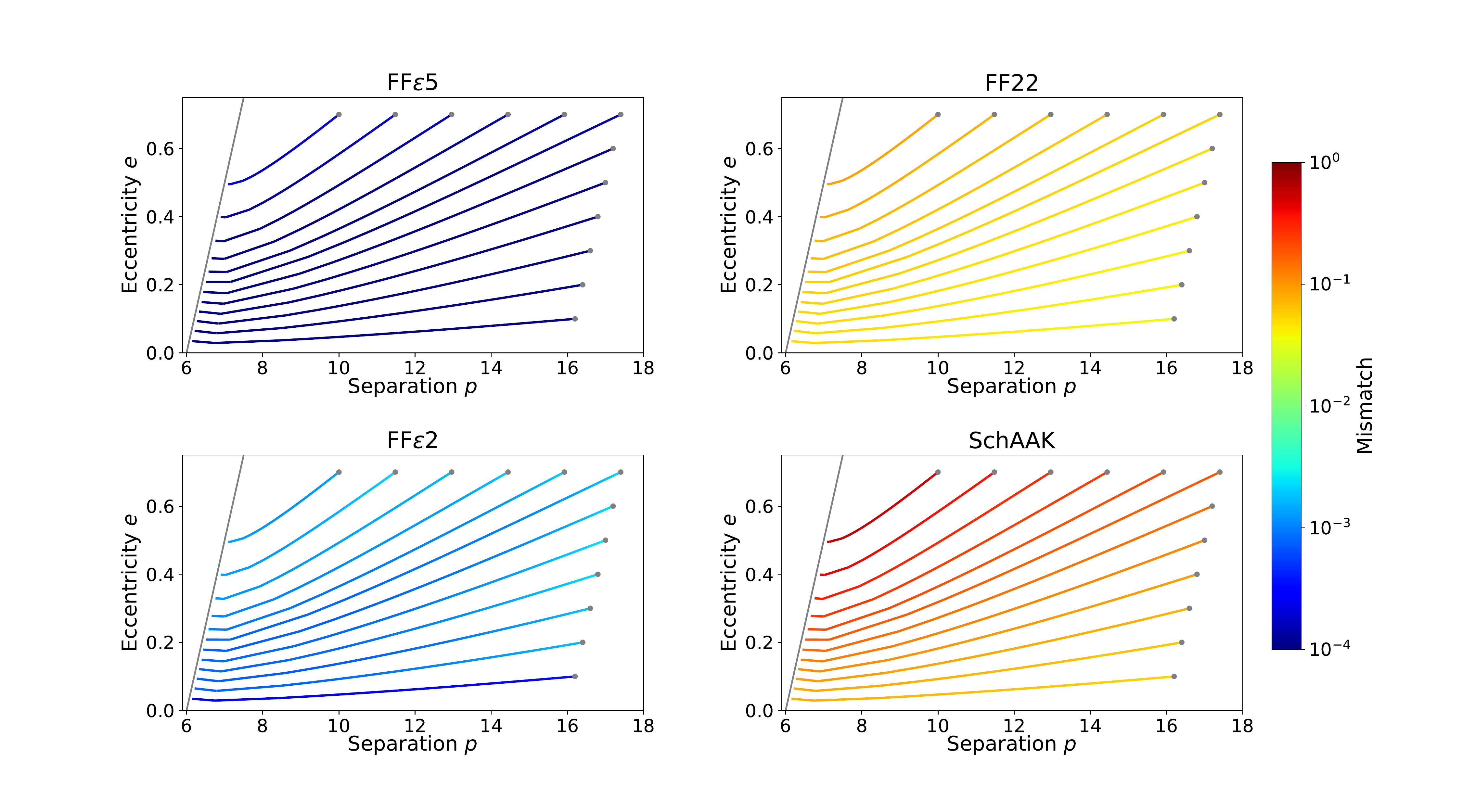}
\caption{For EMRI signals, the harmonic phases and amplitudes create mismatch between comparison waveforms, with the phase strongly dominating. Here, we show the mismatch stemming mainly from amplitude differences between template waveforms (FF$\epsilon$5, FF$\epsilon$2, FF22, SchAAK, which are listed in the title of each plot) and the slow FEW model from \mbox{Section \ref{sec:slow}}. All template waveforms for a given $(p_0, e_0)$ use exactly the same trajectory, but differ in their amplitudes. The trajectories for the slow injection waveforms are slightly more accurate because they use dense-stepping integration, but this small difference does not contribute to the mismatch values shown. The initial values are chosen to outline our parameter space: $(p_0,e_0)\in [(10, 0.7)$, $(11.48, 0.7)$, $(12.96, 0.7)$, $(14.44, 0.7)$, $(15.92, 0.7)$, $(17.4, 0.7)$, $(16.2, 0.1)$, $(16.4, 0.2)$, $(16.6, 0.3)$, $(16.8, 0.4)$, $(17.0, 0.5)$, $(17.2, 0.6)]$. These initial trajectory points are shown with gray dots. The central black hole mass, $M$, is set to $10^6M_\odot$ and the secondary's mass, $\mu$, is determined using FEW utility tools so that the inspiral takes two years to evolve from the initial orbital parameters to the separatrix (shown as the grey line). All other parameters are identical to the injection waveform parameters given at the beginning of Section~\ref{sec:analysis}. The partial mismatch is shown from $t=0$ to $t=t(p,e)$ according to the color bar. As the lines move from the initial gray points to the separatrix, the mismatch is determined over an increasing amount of time. The FF$\epsilon$5 and FF$\epsilon$2 waveforms show strong overlap with the slow and accurate waveform. The FF22 waveform performs marginally and the SchAAK waveform compares poorly, especially at high eccentricity. These differences are visualized in Figure~\ref{fig:waveform_examples}.}\label{fig:mismatch}
\end{center}
\end{figure*}

For this test we analyze 12 EMRI systems with $M=10^6M_\odot$ and with initial parameters that outline the valid parameter space within the FEW model (see the caption of \mbox{Figure \ref{fig:mismatch}} for ($p, e$) values). After setting $(M,p_0,e_0)$ for each EMRI, $\mu$ is adjusted to fix the trajectory duration to 2 years from start position to separatrix. \mbox{Figure \ref{fig:mismatch}} shows the mismatch results for these EMRIs over time for the four waveform models (FF$\epsilon5$, FF$\epsilon2$, FF22, SchAAK) compared against the slow FEW waveform. The curves in the plot are drawn through $(p,e)$-space and show the mismatch from $t=t(p_0,e_0)=0$ to $t=t(p,e)$. Therefore, as $(p,e)$ evolve, the length over which the mismatch is calculated increases until it reaches the full 2 year waveform at the separatrix.

The difference in the various models and settings is clear. FF$\epsilon$5 displays the best mismatch values as expected with a maximum mismatch of $\sim5\times10^{-4}$ in the worst-case waveform scenario with $(p_0,e_0)=(10, 0.7)$. As $(p_0, e_0)$ move towards further separations and lower eccentricities, the mismatch drops to and falls slightly below $10^{-4}$. By adjusting the mode content parameter to $10^{-2}$ (FF$\epsilon$2 waveform), the mismatch increases by a full order of magnitude. Based on \mbox{Figure \ref{fig:mode_content}}, this waveform uses approximately a factor of 10 fewer harmonic modes than the FF$\epsilon$5 waveform. The FF22 waveform has 61 modes in it regardless of the input parameters. At lower (higher) eccentricities, the mode count for the FF22 waveform is more (less) than the FF$\epsilon$2 waveform. This FF22 waveform shows mismatch values between $\sim0.05-0.1$. This is 3 orders of magnitude higher than the base FF$\epsilon5$ waveform. The fourth and final waveform, SchAAK, showed the worst mismatch values as expected due to its semi-relativistic construction. Mismatch values for SchAAK range from $\sim0.05-0.62$. The high mismatch seen for SchAAK at high eccentricities will have strong implications on parameter estimation shown in \mbox{Section \ref{sec:posteriors}}.

\begin{figure*}
\begin{center}
\includegraphics[width=15cm]{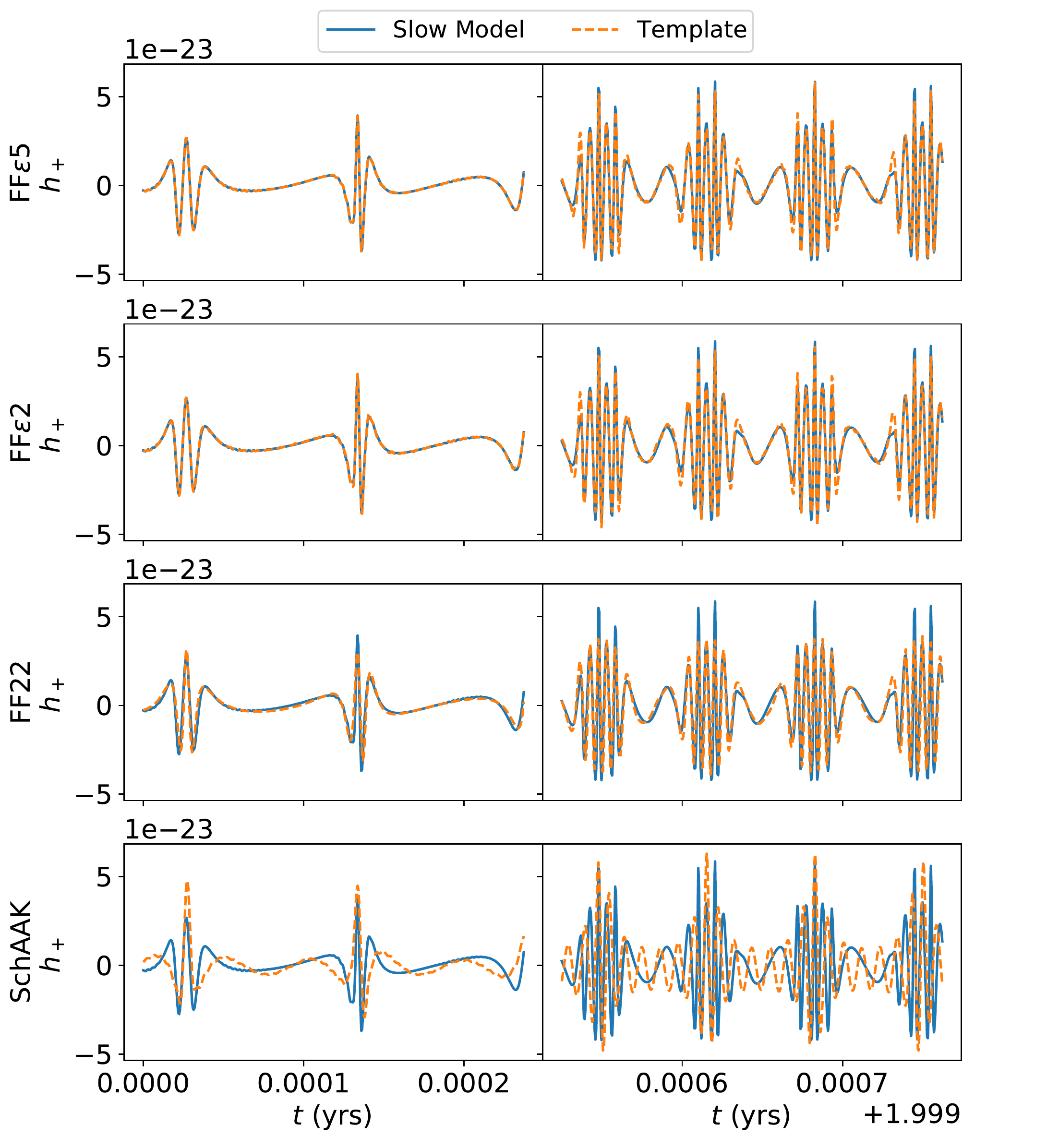}
\caption{Waveforms corresponding to the four waveform models tested in \mbox{Figure \ref{fig:mismatch}} are shown here. Each waveform follows the top-left trajectories in the sub-panels in \mbox{Figure \ref{fig:mismatch}} which have $(p_0, e_0)=(10, 0.7)$ and a final eccentricity $\simeq 0.5$. Each row contains a different fast waveform model labelled along the vertical axis shown with a dashed orange line. These fast models are compared against the slow waveform model appearing as a solid blue line. The left and right plots represent the beginning and end of a two-year waveform. As is expected from the mismatch results, the FF$\epsilon$5 and FF$\epsilon$2 waveforms provide the strongest match to the slow waveform model. The visual difference between these two models is almost indistinguishable. The FF22 quadrupolar waveform begins to have difficulty resolving the higher peaks in the waveform and visually shows a clear separation from the top two rows. The bottom row, with the SchAAK waveform model, clearly reveals the issues related to generating EMRI waveforms with semi-relativistic amplitudes: the overlap is very poor both quantitatively and visually.}\label{fig:waveform_examples}
\end{center}
\end{figure*}

Waveforms comparing each fast model to the slow FEW Schwarzschild eccentric waveform for the $(p_0, e_0)=(10, 0.7)$ trajectory are shown at the beginning and end of the inspiral in Figure \ref{fig:waveform_examples}. This figure visually confirms what we see in the mismatch results. The FF$\epsilon$5 and FF$\epsilon$2 waveforms show strong visual overlap with the slow FEW Schwarzschild eccentric waveform model; these models are also visually almost identical. The quadrupolar relativistic model (FF22) deviates slightly compared to the slow FEW Schwarzschild eccentric waveform model, especially near higher peaks in the strain amplitude. The semi-relativistic SchAAK waveform has difficulty matching the slow FEW Schwarzschild eccentric waveform model at this higher eccentricity which is consistent with the high mismatch values discussed above.

\subsection{Waveform Timing}\label{sec:timing}

Given the sufficient accuracy of a waveform model, the generation time is extremely important to the success of the search for and parameter estimation of gravitational wave sources. For proper Bayesian Markov Chain Monte Carlo (MCMC), greater than $\sim10^6-10^9$ waveform evaluations are necessary for converged posterior distributions. The combination of interpolation methods and acceleration techniques allow the FEW waveforms to be readily used in MCMC studies. 

\begin{figure}
\begin{center}
\includegraphics[scale=0.28]{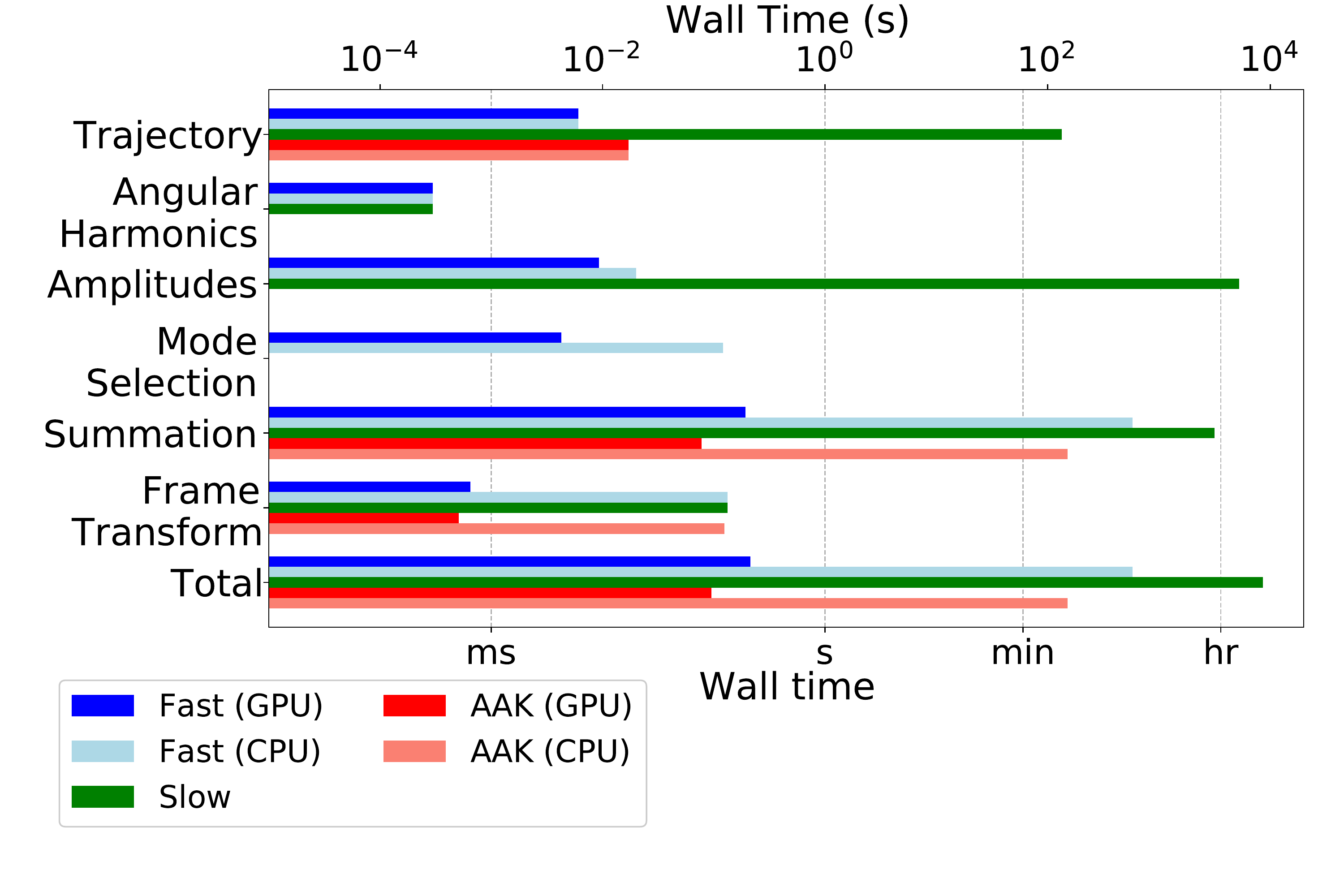}
\caption{Timing benchmarks for various waveforms discussed in this paper. Timing for the trajectory, angular harmonic, amplitude, mode selection, summation, and frame transformation modules is shown from top to bottom, respectively. The sum of those timings is also given. The CPU and GPU versions of the fast FEW waveforms (with $\epsilon=10^{-5}$) (\mbox{Section \ref{sec:schwarz_ecc}}) are shown in light and dark blue, respectively. The slow FEW waveform is shown in green (\mbox{Section \ref{sec:slow}}). The 5PN AAK waveform (\mbox{Section \ref{sec:newAAK}}) is shown in light and dark red, representing the CPU and GPU versions, respectively. $(p_0, e_0)=(10, 0.7)$ for all waveforms, depicting the worst-case in our domain of validity in terms of the number of modes necessary to represent the waveform. The central black hole mass $M$ is set to $10^6M_\odot$ and $\mu$ is chosen to represent a two-year inspiral from the initial $(p_0,e_0)$ to the separatrix. Note the slow waveform does not perform any mode selection. Also, the 5PN AAK waveform does not specifically employ angular harmonic, amplitude, or mode selection modules. The AAK waveform piece (not including the trajectory) is treated in its entirety as a summation module. Timings were computed using a single CPU core on a Xeon Gold 6132 2.60 GHz processor and an NVIDIA V100 GPU. The fast FEW waveform GPU version is $\sim2500\times$ faster than its CPU counterpart.}\label{fig:timing}
\end{center}
\end{figure}

\mbox{Figure \ref{fig:timing}} shows the timing of various pieces of the fast (FF$\epsilon$5) and slow FEW waveforms, with the GPU timing included for the fast waveform. This timing is computed for a two-year waveform in our worst-case configuration (in terms of mode count): $(p,e)=(10, 0.7)$. The CPU computations are run on one core of a Xeon Gold 6132 2.60 GHz processor. The GPU timing is computed with an NVIDIA V100 GPU. 

The trajectory for the fast FEW waveform is the same for the CPU and GPU versions because the trajectory is always computed on the CPU. The large steps in the integrator are a clear advantage of $\sim10^4\times$ faster compared to the dense stepping slow model integrator. Angular harmonics are calculated in the same way for all FEW models, so this timing is always equivalent and faster than 1 ms. The CPU fast model amplitude determination is slightly slower than the GPU implementation because the neural network calculations are performed in parallel on the GPU (CPU computations are tested on a single core but are parallelized when multiple cores are available). However, these computations on both hardware are of order $\sim$10 ms because of the sparse trajectories and efficiency of the neural networks. The slow model amplitudes are determined with the bicubic spline for all modes, a much slower operation: $\sim10^5\times$ slower than the CPU neural network computation. Following the trajectory and amplitude calculations, modes are selected online in the fast FEW models. The sorting operation is the bottleneck here. The GPU sorting algorithm, and, therefore, the mode selection step, is $\sim30\times$ faster on the GPU.

The waveform is then evaluated at the density of the data by interpolating the sparse array information. The summation is the main bottleneck for the fast waveforms, and the piece of waveform generation where the GPU truly separates itself from the CPU implementation. The GPU summation is $\sim3000\times$ more efficient than the CPU summation. The summation bottleneck can be easily understood by examining the number of individual mode computations necessary to build a single waveform. With a conservative data length of $10^6$ points, and each point in the waveform requiring generation of $\sim10^2-10^3$ harmonic modes, a single waveform requires $\sim10^8-10^9$ mode computations. This large number leads to a close similarity in summation times between the fast FEW model CPU version and the slow FEW model with the slow model taking $\sim5\times$ longer. The full waveform timings are dominated by the summation, meaning the comparison of the summation timescales is strongly indicative of the full waveform generation comparison.

Following the summation, the frame transformation is performed. The frame transformation is the same in the slow model and the fast waveform CPU version, but the timing of this step is a small fraction of the summation timescale. However, the GPU maintains its efficiency difference at $\sim200\times$ faster than the CPU-based models for the transformation.  

We also show the timing of the 5PN AAK waveforms (\mbox{Section \ref{sec:newAAK}}) in \mbox{Figure \ref{fig:timing}}. To be clear, this subsection is the only part in our results where we test the full 5PN AAK waveform including $(a, Y_0, \Phi_{\theta, 0})=(0.5, 0.77, 0.0)$. The 5PN trajectory is slightly longer than the relativistic adiabatic Schwarzschild trajectory because of the long analytic formulae necessary for 5PN calculations (rather than a basic interpolation in the relativistic trajectory) and the necessity of computing two more derivatives ($\dot{Y}, \Omega_{\theta}$) in the higher dimensional parameter space. 

The AAK piece of the waveform is implemented as a special summation module. There are no specific angular harmonic, amplitude, or mode selection computations in the AAK waveform, so those parts of the computation are excluded. The AAK summation shows a similar timescale ratio to the relativistic waveforms between its GPU and CPU versions. However, the AAK summations are generally $\sim2\times$ faster than the worst-case relativistic waveform. Since the frame transformation computations are the same as the relativistic waveform, the overall waveforms show similar timing comparisons to the summation timescales. However, it is interesting to note that at higher $\epsilon\gtrsim10^{-2}$, where less harmonic modes are required, the fast relativistic waveform can become faster than the 5PN AAK. This is simply explained since the AAK summation involves a longer sequence of computations for each harmonic mode at each time point. The fast waveform summation, on the other hand, is much simpler, employing straight-forward interpolation and complex multiplication to determine each individual mode value.

\subsection{Intrinsic Posterior Analysis}\label{sec:posteriors}

It is helpful to transform the mismatch information into actual metrics for data analysis. Here, we will focus on determining measurement precision and bias for the waveforms analyzed in this section. Rather than using the typical Fisher matrix \cite[e.g.][]{Vallisneri:2007ev} and Cutler-Vallisneri bias \cite{Cutler:2007mi} calculations, we leverage the efficiency of the GPU implementations to generate full Bayesian posteriors from basic standard parameter estimation runs.

The focus of this paper is on the new waveform models. Therefore, we only briefly describe our parameter estimation techniques and settings; we refer the interested reader to cited papers for more information. 

The posterior distribution represents the product of the prior distribution and likelihood distribution. The prior and likelihood terms are normalized by an intractable normalization factor. Markov Chain Monte Carlo (MCMC) methods sample from the posterior to build posterior distributions from the density of samples. Since all samples are normalized by the same value, it is not necessary to calculate this normalization factor for accurate posterior results. We employ uniform priors spanning large fractions of the parameter space and the typical gravitational wave likelihood as defined above.

For sampling, we use a slightly modified version of the \texttt{emcee} \cite{emcee} package that helps account for periodic parameters. Initial starting points for runs were sampled from a multivariate normal distribution with mean equal to the injection parameters and covariance matrix determined with Fisher methods for the template model (not the injection model). The burn-in allows these walkers formed tightly around the true point to spread out and properly cover the posterior. The proposal used is the default Affine-Invariant or Stretch proposal \cite{Goodman2010}. We used 32 walkers in each run. The average autocorrelation time, $\hat{\tau}$, is determined across chains using \cite{autocorrelation}. The burn-in used was chosen to be $2\hat{\tau}_\text{max}$. Chains were thinned by a factor of $\frac{1}{2}\hat{\tau}_\text{min}$ \cite{autocorrelation}, where $\min$ and $\max$ indicate the minimum and maximum autocorrelation times across the $D=6$ dimensional parameter space. The sampler was run until chains were longer than $50\hat{\tau}$, providing an effective sample size (ESS) of $\sim3000$ for each case tested. The particular sampling method used will keep walkers from moving far from the main mode of the posterior. Therefore, our study does not comment on any posterior phenomena away from the main posterior mode. 

Given the FF$\epsilon5$ waveform shows excellent agreement with the slow waveform, we inject waveforms with this model. Signals are injected without noise in order to understand the pure systematic bias resulting from the different waveform models. All injections are scaled to a chosen SNR of 30 (this effectively means scaling the distance) over 2 years of inspiral. The SNR is chosen to represent an astrophysically motivated source \cite[e.g.][]{Babak2017}. All sources analyzed are located at a sensible distance with most distances at $\sim1$ Gpc. The closest source is located at $\sim0.2$ Gpc. Once injected, we perform independent parameter estimation runs for all parameter sets using FF$\epsilon5$, FF$\epsilon2$, FF22, and SchAAK waveforms as templates.

The parameters tested were only the intrinsic parameters: $\{M, \mu, p_0, e_0, \Phi_{\varphi,0}, \Phi_{r,0}\}$. The initial phases $(\Phi_{\varphi,0}, \Phi_{r,0})$ used for the injection were set to (3.23, 4.72). $(M, \mu, e_0)$ were chosen from a $3\times3\times4$ grid. We choose $M$ from the set $\{3\times10^5M_\odot, 10^6M_\odot, 3\times10^6M_\odot\}$; we choose $\mu$ from the set $\{ 3 M_\odot, 10 M_\odot, 30 M_\odot \}$; and we choose $e$ from the set $\{ 0.1, 0.3, 0.5, 0.7 \}$. After ($M, \mu, e_0$) are set, $p_0$ is determined to be $\min(p_0(t=2 \text{yrs}), 16.0 + 2e)$. $p_0(t=2 \text{yrs})$ indicates the $p_0$ value at 2 years before reaching the separatrix given the chosen ($M, \mu, e_0$).

This parameter and waveform grid amounts to 144 separate posteriors.  We remove from this four sets of parameters, corresponding to $(M, \mu, e_0)$ = $(3\times10^6, 3.0, 0.7)$, $(3\times10^6, 10, 0.7)$, $(3\times10^6, 30.0, 0.7)$, and $(10^6, 3.0, 0.7)$.  In these cases, the parameters correspond to $p_0(t=2 \text{yrs})<9.9$, which is outside of our domain of validity at $e=0.7$.  With the GPU acceleration, most runs were $\sim1-2$ hours in duration. Posterior runs with multimodality were longer at $\sim6-10$ hours due to larger autocorrelation times. 

\begin{figure*}
\begin{center}
\includegraphics[width=0.87\paperwidth]{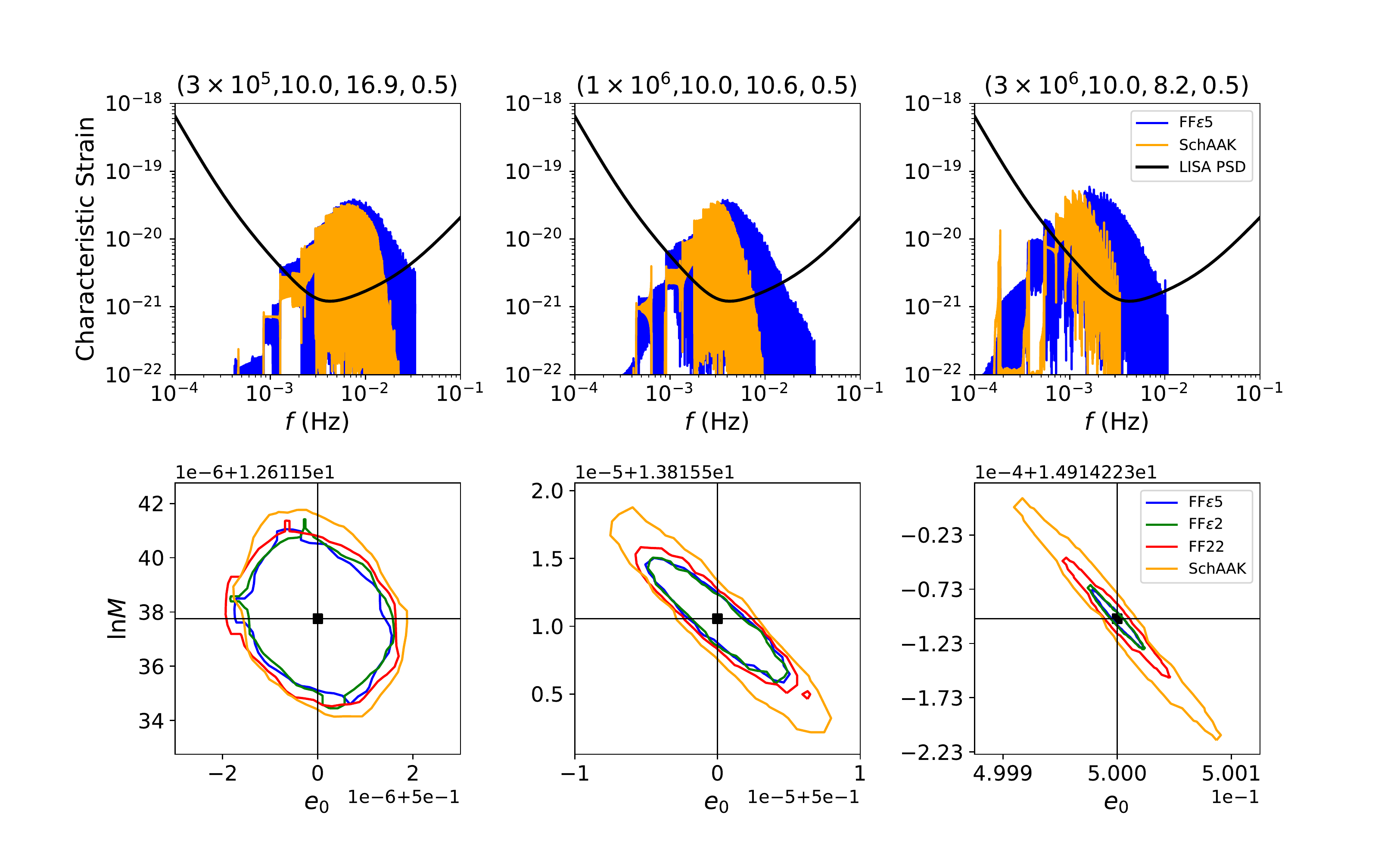}
\caption{Characteristic strain (top row) and two-dimensional $3\sigma$ posteriors in the $e_0-\ln{M}$ plane (bottom row) for three different binary configurations. Above each top row plot the parameters are listed as $(M, \mu, p_0, e_0)$ and each plot is made with $\mu=10$ and $e_0=0.5$. From left to right, injections have $M$ values of $3\times10^5$, $1\times10^6$, and $3\times10^6$. Within each bottom row plot, the FF$\epsilon5$, FF$\epsilon2$, FF22, and SchAAK models are shown in blue, green, red, and orange, respectively. Black horizontal and veritcal lines indicate the location of the injection point. The top plots illustrate the spectral difference between FF$\epsilon5$ (blue) and SchAAK (orange) waveforms at the injection point compared against the LISA sensitivity curve. For completeness, we also tested all parameter sets in this plot again with the noise contribution from the Galactic foreground. It did not change the results by any significant amount.}\label{fig:concentric}
\end{center}
\end{figure*}

When comparing models to each other, a clear behavior is observed. \mbox{Figure \ref{fig:concentric}} shows one example comparing all waveform models tested at various $M$ with $(\mu, e_0)=(10, 0.5)$. The posterior distributions plotted in the bottom row of the figure are the two-dimensional marginalized posterior distributions in the $e_0$-ln$M$ plane. The contours shown correspond to the $3\sigma$ value for a two-dimensional Gaussian distribution (described as the $3\sigma$ contour below). For a given orbital configuration the mass $M$ sets the frequency band over which an EMRI radiates. The relation of the frequencies in the radiation to the noise curve of LISA will effect the behavior of the posterior distributions. At $M=3\times10^5M_\odot$, the EMRI radiates most of its power between $\sim3-40$ mHz, which is considered the high-frequency end of the LISA noise curve. Here, the noise increases as the frequency is increased. This means that higher frequency modes will experience more noise suppression than lower frequency modes. Since the FF22 and SchAAK waveforms are built from only quadrupolar radiation, these waveforms will contain less signal at the higher frequency end when compared to FF$\epsilon5$ and FF$\epsilon2$ waveforms. This is especially true with higher eccentricity. This concept is illustrated in the top row of \mbox{Figure \ref{fig:concentric}}. The top row of plots shows comparisons of the Fourier Transform ($\tilde{h}(f)$) of the time domain signals built with the FF$\epsilon5$  and SchAAK models in the characteristic strain representation: $h_c^2=f^2|\tilde{h}(f)|^2$ \cite{Moore2015Sensitivity}. With the higher frequency modes suppressed by the noise, we find similar posteriors for all models as expected. The differing models show effectively no bias on this source with close, concentric posterior distributions of the FF$\epsilon5$, FF$\epsilon2$, FF22, and SchAAK models in that order from middle to outer ellipse. To ensure our analysis was appropriate, since we do not include the Galactic foreground noise expected for LISA, we repeated the tests shown in \mbox{Figure \ref{fig:concentric}} while \textit{including} the Galactic foreground and found no significant changes to the results that follow. 

As the mass of the MBH is increased to $M=10^6M_\odot$, the waveforms shift towards the center of the LISA band, peaking directly over the most sensitive portion at $\sim4$ mHz. Here, both higher and lower frequency modes are similarly suppressed. In this case, there is no ``enhanced weighting'' of the lower frequency modes. This leads to a stronger difference in posterior distributions with the FF22 and SchAAK models spreading further away from the injection point compared to the FF$\epsilon5$ and FF$\epsilon2$ waveforms.

The highest mass case with $M=3\times10^6M_\odot$ shifts the radiation frequencies to the lower frequency end of the LISA sensitivity band. At this end, the noise decreases as frequency is increased. Therefore, here, the lower frequency modes are now further suppressed by the noise compared to the higher frequency. This results in a further widening of the posteriors from the quadrupolar waveform templates. 

For all masses tested, the FF$\epsilon2$ model strongly mirrors the behavior of the FF$\epsilon$5 model. This is an initial indication that higher mode counts may not be necessary for accurate parameter estimation of EMRI sources. However, what is necessary is that the modes be intelligently chosen to represent the relativistic behavior of the EMRI system rather than simply fixing a predetermined set of modes at low order in $l$.

\begin{figure*}
\begin{center}
\includegraphics[scale=0.53]{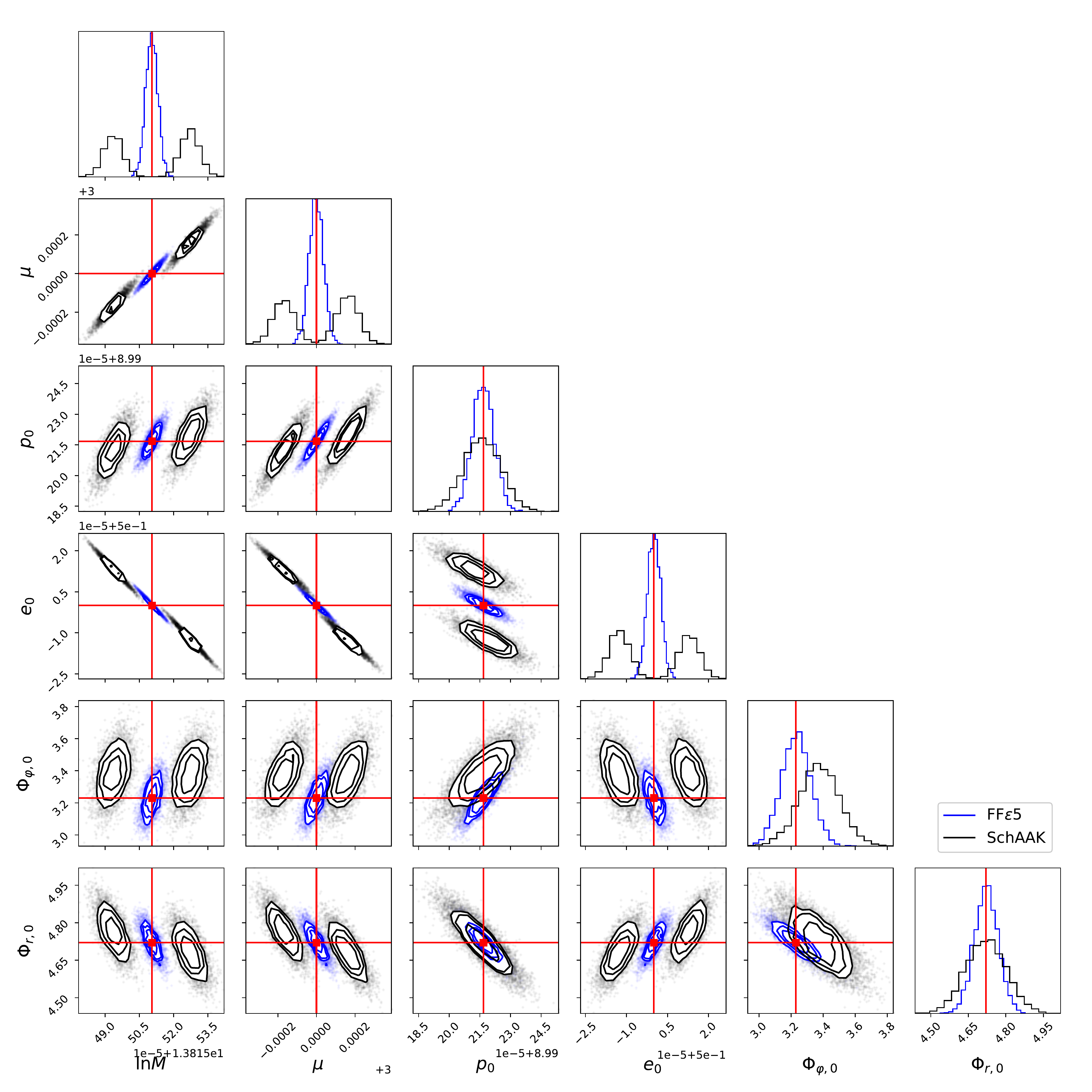}
\caption{Corner plot showing one- and two-dimensional marginalized posteriors for an injection with $(M, \mu, p_0, e_0, \Phi_{\varphi,0}, \Phi_{r,0})=(10^6M_\odot, 3M_\odot, 8.99, 0.5, 3.23, 4.72)$. The histograms and contours for the $\epsilon=10^{-5}$ fast FEW (FF$\epsilon5$) and Schwarzschild AAK (SchAAK) template waveforms are shown in blue and black, respectively. The true injection parameters are denoted with the red vertical and horizontal lines. Note the sampling methods used concentrated near the true value. The samplers are not be able to access secondary modes with a strong separation from the injection point. Due to the inaccuracy of the semi-relativistic amplitudes in the SchAAK waveform, a strong bias and multimodal behavior are observed when fitting a SchAAK waveform template against an FF$\epsilon$5 relativistic injection. The consequences of this behavior are further discussed in Section~\ref{sec:discuss}.}\label{fig:posterior1}
\end{center}
\end{figure*}

\mbox{Figure \ref{fig:posterior1}} shows an example of a full posterior with injection parameters $(M, \mu, e_0)=(10^{6}M_\odot, 3 M_\odot, 0.5$). The blue and black posteriors represent the FF$\epsilon5$ and SchAAK template models, respectively. Therefore, the blue posterior shows a direct test against the injection. All blue output distributions are unimodal, Gaussian, and centered around the true point with no bias, as expected. The black posterior shows what can happen when trying to fit a relativistic injection with a semi-relativistic template model. There is a strong bias resulting in the true injection point positioned far outside the 3$\sigma$ posterior contours. The posterior is multimodal, displaying one-dimensional multimodality in the $\ln M,\ \mu$ and $e_0$ parameters. In general, the gravitational wave likelihood function is negatively affected by mismatch in the phase and the amplitude of a signal; but, the relative effect of each can vary, with the phase dominating. Since the phasing in both models shown is identical, we expect and observe the posterior weight to be located close to the injection point. However, the amplitude-based mismatch for the SchAAK waveform against the relativistic injection across our domain of validity is of order $\sim0.1-0.6$ (see \mbox{Figure \ref{fig:mismatch}}). At higher eccentricity, the mismatch approaches $\sim0.62$. This amplitude mismatch can produce points in parameter space with distinct differences in the intrinsic parameters from the injection that show better alignment with the injected waveform. In fact, in the case shown in \mbox{Figure \ref{fig:posterior1}}, there is effectively no posterior weight at the injection point. This results in posterior modes at slightly higher (lower) ln$M$, higher (lower) $\mu$, and lower (higher) $e_0$, which can be observed in the parameter correlations shown in the two-dimensional marginalized posteriors. 

The SchAAK template at the injection and best-fit (maximum likelihood) parameters are visually similar, except for near the separatrix where the best-fit waveform begins to dephase from the injection. The SchAAK waveform at the injection does not dephase compared to the FF$\epsilon$5 injection because their trajectories are identical. The best-fit waveform has different intrinsic parameters meaning its trajectory and, therefore, the time at which it reaches the separatrix is different from the injection causing this dephasing. Consequently, it is interesting that despite the slow dephasing of slightly different trajectories, the best-fit parameters still provide a better match to the relativistic injection waveform. The best-fit parameters produce a mismatch of 0.50, a $\ln \mathcal{L}\approx-313$ ($\ln \mathcal{L}=0$ for the injection against itself), and an extraction SNR of $\sim22.3$ (SNR of the injection is 30). A SchAAK waveform generated at the injection parameters has an associated mismatch of 0.53,   $\ln \mathcal{L}\approx-347$, and an SNR of $\sim21.8$. Therefore, $\Delta \ln \mathcal{L}\approx34$, indicating the injection point is not expected to be within the explored posterior (the minimum $\ln \mathcal{L}$ value included in the posterior is $\sim-330$). 

\begin{figure*}
\begin{center}
\includegraphics[scale=0.48]{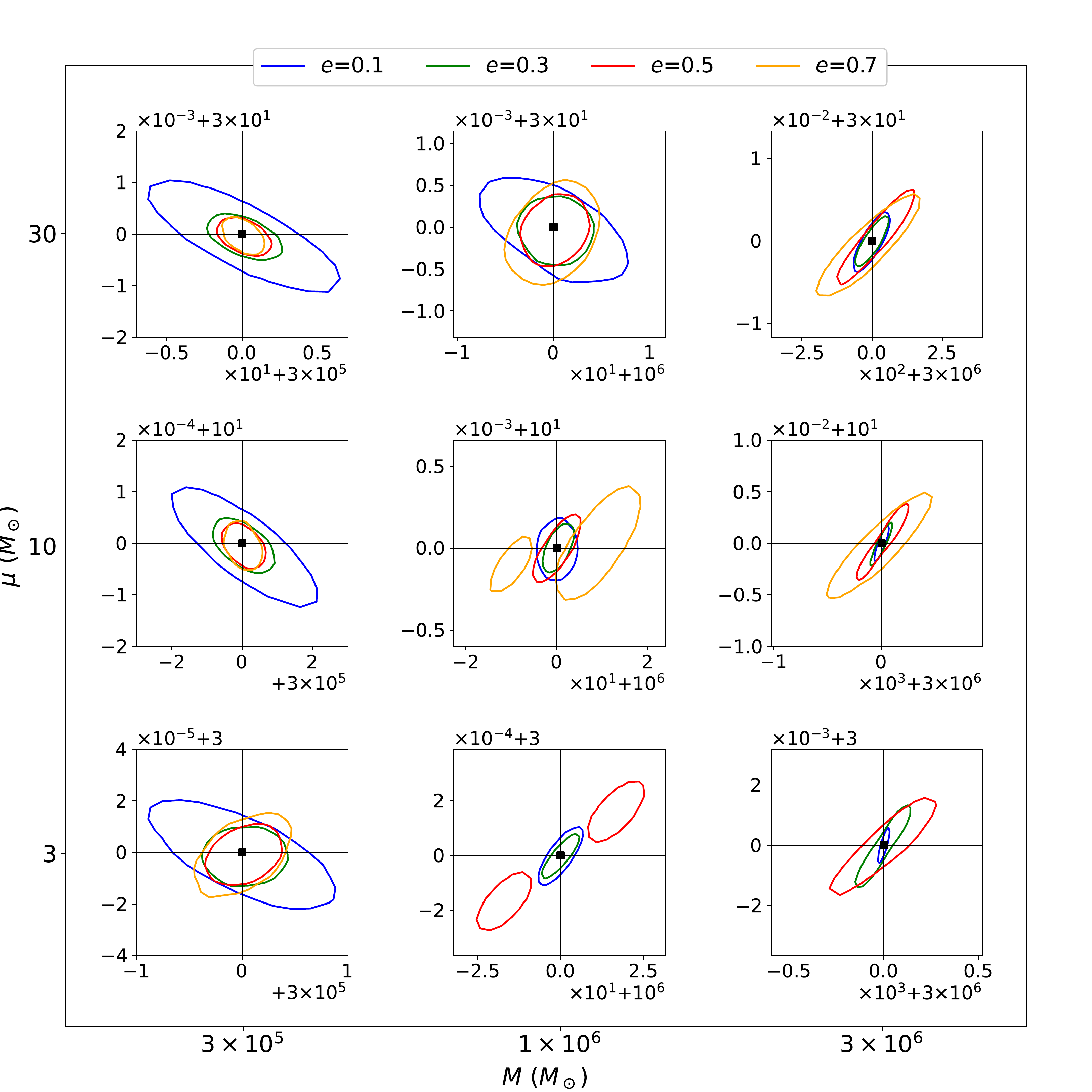}
\caption{Two-dimensional $3\sigma$ posteriors for $M$ and $\mu$ parameters are shown for all injections tested using the SchAAK waveform as the fitting template to the FF$\epsilon$5 relativistic injection. Each two-dimensional posterior is arranged horizontally and vertically according to the $M$ and $\mu$ injection values, respectively. Within each subplot, posteriors are shown for every injection value of eccentricity tested with eccentricities of 0.1, 0.3, 0.5, and 0.7 shown in blue, green, red, and orange, respectively. Vertical and horizontal black lines indicate the true values of the injection. Please note injections with $(\mu, e)=(3, 0.7)$ and $M\in[1\times10^6, 3\times10^6]$ do not fall within our domain of validity and, therefore, are not shown.}\label{fig:combined1}
\end{center}
\end{figure*}

\begin{figure}
\begin{center}
\includegraphics[scale=0.55]{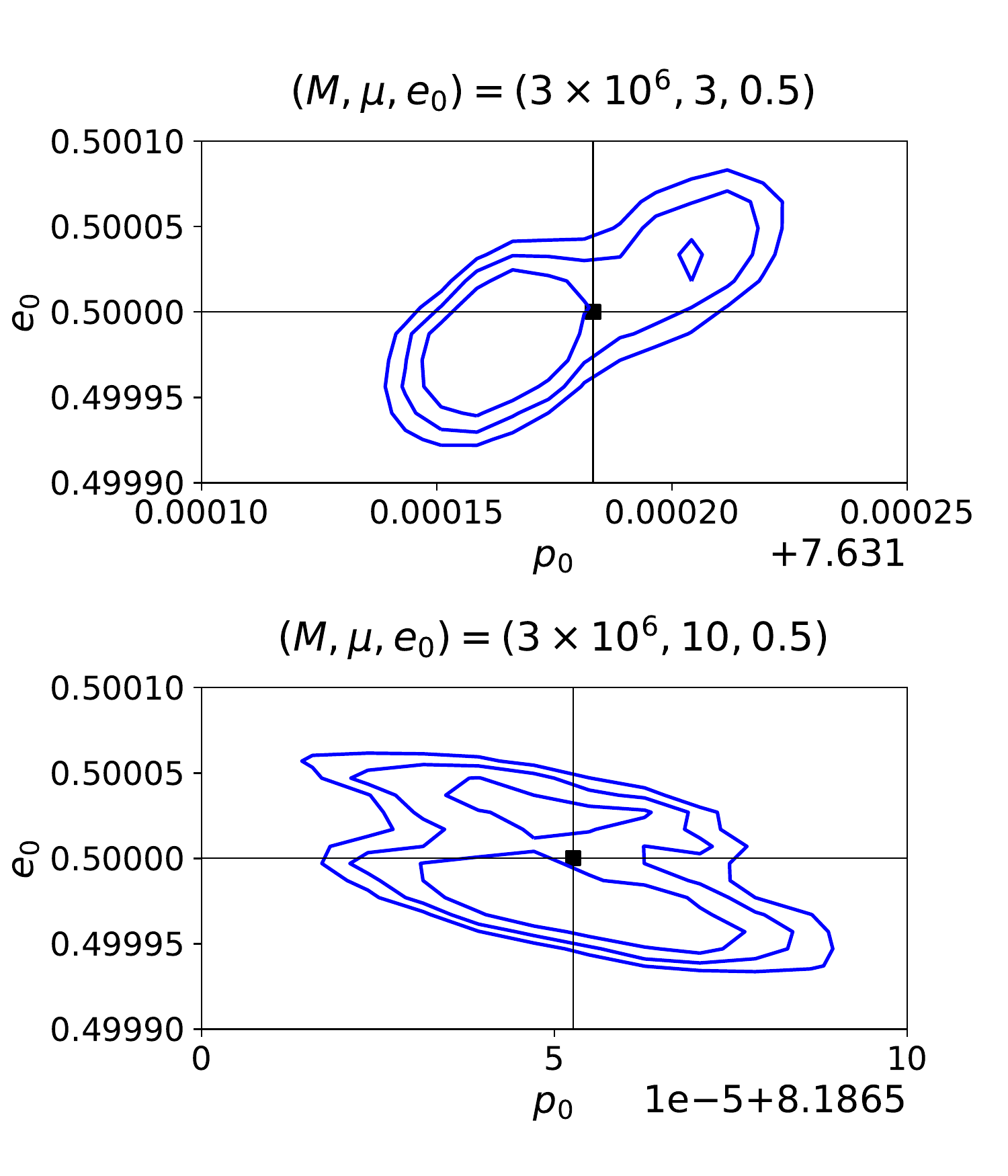}
\caption{Multimodality in the $p_0$-$e_0$ plane is shown above for injections with $(M, e_0)=(3\times10^6, 0.5$) and $\mu$ values of 3 and 10 in the top and bottom plots, respectively. In each plot, the 1$\sigma$, 2$\sigma$, and 3$\sigma$ contours are shown for the SchAAK template waveform. Black horizontal and vertical lines give the true injection values. The two cases shown here exhibit only a slight bias with the true injection point contained within the $2\sigma$ contour.}\label{fig:other_multi}
\end{center}
\end{figure}

Similar multimodal behavior is observed for the SchAAK template with $(M, \mu, e_0)=(10^6,10, 0.7)$. This is shown in \mbox{Figure \ref{fig:combined1}}. In this figure, we show, for the SchAAK template, a  3$\times$3 grid of plots arranged according to $M$ and $\mu$ of the injection. Within each of these nine plots, the two-dimensional $M-\mu$ 3$\sigma$ posterior is shown at all four eccentricity injection values. This figure shows the multimodal and bias behavior across all injections tested. However, it must be noted there is multimodal behavior for $(M,\mu, e_0)\in[(3\times10^6, 3, 0.5), (3\times10^6, 10, 0.5)]$, which is shown for $(p_0,e_0)$ parameters in \mbox{Figure \ref{fig:other_multi}} (other parameter pairs do show multimodality, but only $(p_0,e_0)$ is shown to be succinct). In the $(M,\mu, e_0)=(3\times10^6, 10, 0.5)$ case, multimodality is observed in $M$, $\mu$, and $e_0$, but the the bias is not as strong as the first two cases discussed with two posterior modes observed at the level of the 1$\sigma$ contour with the $2\sigma$ contour surrounding the two $1\sigma$ modes. For $(M,\mu, e_0)=(3\times10^6, 3, 0.5)$, the multimodality is found only in the $p_0$ parameter and the bias is small: the injection point does fall within the 2$\sigma$ contour. 

It is hard to predict exactly when and how the bias and multimodal behavior will manifest. From our results, we expect it to occur at higher $M$, lower $\mu$, and higher $e_0$. This manifests from the spread in harmonic modes towards lower frequencies, similar to the previous discussion around \mbox{Figure \ref{fig:concentric}}. However, these behavioral expectations are not fixed rules as the case with $(M, \mu, e_0)=(3\times10^6, 10, 0.7)$ shows no visual bias or multimodality, indicating it is not easy to predict exactly when or how this behavior will occur. This uncertainty is due to the inherent difference between the SchAAK waveform manifold and the relativistic waveform manifold, making it hard to directly understand waveform comparisons at each location in parameter space. However, it must be noted that no posteriors generated with the FF22 quadrupolar template exhibit any multimodal or biased behavior.  The key point here is that the semi-relativistic amplitudes of the SchAAK model can result in a reasonable extraction of posterior distributions, but includes the risk of producing extraneous posterior structure and potential bias. 

\section{Discussion and Future Outlook}\label{sec:discuss}

The modular FEW framework is designed to facilitate updates to the waveform models. Future work will focus on extending the fully relativistic models to Kerr spacetime, improving the phase accuracy of the waveforms, and computing the waveforms in domains other than the time-domain. Furthermore, the speed of the FEW model allows for efficient exploration of the LISA data analysis problem. These tasks can be carried out in parallel with each other; we discuss each task in the subsections below.

\subsection{Extending the fully relativistic waveform amplitude model to generic Kerr inspirals}\label{sec:genKerr}

From Schwarzschild inspirals it is natural to first extend the amplitude model to eccentric, equatorial or spherical orbits in Kerr spacetime as this only increases the dimension of the parameter space by one. The model can then be extended to the four dimensional parameter space of generic (eccentric and inclined) inspirals into a Kerr black hole. Frequency domain Teukolsky codes exist which can rapidly compute the waveform amplitudes across these expanded parameter spaces \cite{Hughes2021EMRI_FD}; data sets providing waveform data for spherical and equatorial eccentric orbits for many spins can be found in the Toolkit \cite{BHPToolkit}, and work is in progress to generate similar data for generic orbits. We expect that the RomanNet method presented in Section~\ref{sec:RomanNet} will extend well to efficiently interpolate the amplitudes across higher dimensions.

To model Kerr inspirals the spin-weighted spherical harmonics, $_{-2}Y_{lm}$, will also need to be adjusted to the spin-weighted spheroidal harmonics, $S_{lmkn}(t, \theta)e^{im\phi}$. The spheroidal harmonics add both an additional two harmonic indices and time-dependence compared to the spherical harmonics.  Both of these complications can be modeled by using the fact that the spheroidal harmonics can be expanded in spherical harmonics:
\begin{equation}
    S_{lmkn}(t,\theta)e^{im\phi} = \sum_{j = l_{\rm min}}^\infty b^j_{lmkn}(t) {_{-2}Y}_{jm}(\theta,\phi)\;,
    \label{eq:spheroidexpand}
\end{equation}
where $l_{\rm min} = {\rm max}(2, |m|)$.  Appendix A of Ref.\ \cite{Hughes2000} discusses this further, and describes how to compute the spheroidal-spherical mixing coefficients $b^j_{lmkn}(t)$.  Over most of parameter space, we find the approximate scaling $b^j_{lmkn}(t) \sim (a\omega_{mkn}(t))^{|j - l|}$.  The expansion coefficients thus peak at $j = l$, and fall off as powers of $a \omega_{mkn}(t)$ away from this peak.  By knowing these coefficients over the inspiral, we can project the waveform amplitudes onto a spherical harmonic basis, rewriting Eq.\ (\ref{eq:main_wave}) as
\begin{equation}\label{eq:main_wave_spherical}
    h = \frac{\mu}{d_L}\sum_{lmkn} {\cal A}_{lmkn}(t) {_{-2}}Y_{lm}(\theta,\phi)e^{-i\Phi_{mkn}(t)}.
\end{equation}
It is worth noting that, especially in the strong field and for large $a$, there are many orbits which have $a\omega_{mkn} > 1$.  In such cases, the approximate scaling of $b^j_{lmkn}$ does not hold.  We nonetheless find even then that the expansion (\ref{eq:spheroidexpand}) converges with a finite number of terms, making it possible to implement Eq.~\eqref{eq:main_wave_spherical}.  See Appendix A of Ref.\ \cite{Yunes:2010zj} for further discussion, and an explicit relation between the spheroidal amplitudes $A_{lmkn}$ and the spherical amplitudes ${\cal A}_{lmkn}$. Once the data are organized in this way, the summation module can be easily extended to generic Kerr.

For Kerr inspirals the amplitude determination and mode sorting is expected to take longer relative to Schwarzschild eccentric computations due to the increase in the number of harmonic modes. However, we expect these computations to be faster than the final waveform summation, as is the case in the current implementation. Like the amplitude and mode sorting modules, the waveform summation is expected to increase in duration as the number of modes required in generic Kerr substantially increases. This expectation indicates that a deeper analysis on how to best include amplitude information will be useful.

A key component of online waveform generation efficiency is the ability to select harmonic modes based on their contribution to the total power. Currently, this involves both generating amplitudes for and sorting every harmonic mode. This operation is expensive; however, the waveform summation bottleneck is worse and would be much worse if every waveform was produced with \textit{every} mode. As the speed of the waveform summation is improved, mode selection will become the bottleneck. Improved methods of online mode selection will be required. We expect a useful form of this would be an effective precomputed ``mask'' applied prior to generating the amplitudes. 

\subsection{Improvements to the phase accuracy of the models}

Whereas the waveform amplitudes only need to be known to adiabatic order \cite{Miller2020_twotimescale} to enable the full potential of EMRI science, the waveform phase must be computed to post-adiabatic order \cite{Hinderer2008twotimescale}. This presents three challenges: (i) the adiabatic contributions to the phase need to be interpolated to a precision better than $1/q$ ($q=\mu/M)$ across the, up to, four dimensional parameter space \cite{Osburn:2015duj}, (ii) the inclusion of orbital resonances, and (iii) post-adiabatic corrections must be computed. The latter includes conservative corrections to the orbital dynamics \cite{Barack:2010tm,van_de_Meent_2018}, second-order in the mass ratio corrections \cite{Pound20202ndOrder, Warburton:2021kwk}, and corrections due to the spin of the secondary \cite{Harms:2015ixa,Warburton:2017sxk,Akcay:2019bvk,Piovano:2020zin,Skoupy2021spinningtestbody}. As all of these contribute $\mathcal{O}(q^0)$ radians to the waveform phase they do not need to be interpolated as accurately as the adiabatic contributions \cite{Osburn:2015duj}. Substantial work is required to complete these calculations and to sample the parameter space efficiently, but, as these results become available, they can be seamlessly incorporated into FEW.

One challenge with the post-adiabatic phase corrections is that some of them introduce oscillations on the orbital timescale. This can drastically slow down the numerical integration of the phase trajectory from seconds to minutes or hours depending on the mass ratio \cite{Osburn:2015duj}. Fortunately this can be overcome with the use of schemes that average over the short orbital timescale while capturing the correct long term phase evolution of the binary \cite{van_de_Meent_2018b,Miller2020_twotimescale}. With these implemented the calculation of the inspiral trajectory takes milliseconds \cite{van_de_Meent_2018b}. When using averaging methods such as those described in Ref.~\cite{van_de_Meent_2018}, a final phase refinement step on the orbital timescale may be required. We have begun developing this computation and our early findings show the timing of this calculation is a small fraction of the overall waveform summation speed.

Finally, our models need to include the effects of orbital resonances \cite{Flanagan2012_resonance,Ruangsri:2013hra,Brink2015_resonance}. These produce a short-lived kick to the orbital phase whenever the polar and radial frequencies of the orbit are in a low-integer ratio. Precisely modelling the resonances requires knowledge of the post-adiabatic corrections to the phase but approximate models  \cite[e.g.][]{speri2021assessing} can already be incorporated into the FEW framework while the full resonant model is developed.

\subsection{Signal and Data Analysis}\label{sec:data_analysis}

Two immediate benefits of the FEW framework are the waveform acceleration and useful set of flexible tools and modules. The GPU acceleration allows us to test search and parameter estimation algorithms in a tractable amount of time. This is shown in force above by the ability to run $\sim100$ independent posterior analyses on the timescale of a week. Prior to and following these analyses we also used the FEW tools and modules to prepare information for sampling runs or to analyze and understand their outputs.

Our investigations of posterior distributions show important information for future tests of EMRI search and parameter estimation. The main finding was expected: the phasing of the EMRI is the leading order effect on the posterior distributions when comparing models. Since this phasing was the exact same for all models, it is clear the differing amplitudes generally cause a small or negligible bias. Certain cases at higher eccentricity, higher $M$, and lower $\mu$ show a variety of multimodal behaviors when analyzing the relativistic injection with a semi-relativistic template. The lack or presence of multimodality and bias across the various models has implications for parameter estimation:  \textit{using a kludge model may artificially inflate the number of posterior modes making both search and parameter estimation more difficult}. Conversely, settings in FEW that analyze fewer harmonic modes show that the sacrifice in accuracy (and remaining lack of close multimodality) is worth the improved speed.

A transition to plunge and inclusion of the merger-ringdown is still required to more accurately model these waveforms. It is true that the SNR contained near and after plunge is small compared to the overall SNR for an EMRI since the signal SNR is effectively linearly increasing with time. However, it would still be useful to model this piece of the waveform to maxmimize SNR and have a fuller picture of the morphology of the EMRI signal in the strong-field regime. Additionally, we anticipate that the FEW framework will be extend to model intermediate mass ratio inspirals ($\mu/M\sim10^{-2}-10^{-4}$) by the inclusion of post-adiabatic corrections \cite{Warburton:2021kwk}. In this regime, the merger-ringdown will play a more important role in characterizing the signal.

The merger-ringdown cannot be directly added into the existing FEW framework because, near plunge, the approximation of modeling the waveform as a sequence of bound orbits breaks down. This piece of the waveform would be implemented as an independent module that will be attached to the original waveforms. For more information, see \cite{Hughes:2019zmt, Apte:2019txp, Lim:2019xrb}. 

Future gravitational wave data analysis may take place in domains other than the time domain. We plan to expand FEW beyond computing time domain waveforms to also compute waveforms in the frequency, time-frequency, and wavelet domains \cite{Cornish2020Wavelet}. FEW is designed to handle this change. We expect waveforms computed directly in the frequency domain \cite{Hughes2021EMRI_FD} to be more efficient to generate. Waveforms built in the time domain require the evaluation of all contributing modes at \textit{every} time point. This means the waveform is built with $\sim n_t\times n_{lmkn}$ separate mode evaluations. When building the frequency-domain waveform, modes need only be evaluated at their contributing frequencies. Since each harmonic mode evolves in frequency over only a small subset of bins in the Fourier transform, the number of 
harmonics associated with each frequency bin will vary. This indicates less total mode evaluations are necessary, therefore, improving the overall speed of waveform generation. 
The time-frequency and wavelet domains will also play a crucial role in gravitational wave analysis due to non-stationary effects \cite[e.g.][]{Cornish2020Wavelet}. EMRIs are long-lived signals, meaning the non-stationary effects may strongly bias measurements. Studies of this nature for EMRIs are topics of future work. EMRIs may also uniquely benefit from time-frequency-type methods due to their rich harmonic structure. Therefore, these domains will be implemented in FEW as soon as they are available. 

As new domains are implemented, it will generally mean slight changes to the FEW summation modules and the information supplied to those modules. Therefore, we expect only small changes to the code. However, after these methods are added, the bottlenecks in the code may change to areas other than the waveform summation. This would require further innovation in specific operations where less effort has been concentrated so far.

For a full analysis of EMRI signals with LISA, effort will be needed to create efficient implementations for the LISA response. The LISA response is time-dependent as the detector orientation to the source rotates throughout its orbit. Constructing the LISA response accurately in the variety of domains will be paramount to maintaining analysis speed and accuracy.

\section{Conclusion}

We presented the FastEMRIWaveforms framework and package in detail, expanding on the original {\it Physical Review Letter} that introduced the FastEMRIWaveforms Schwarzschild eccentric waveform template for data analysis \cite{Chua2020RapidGenLetter}. The framework is built in a highly modular structure that contains stand-alone modules that are combined into full waveform models. These individual modules provide great flexibility for future applications where methods or physical information is further developed. The user interface for this package is in \texttt{Python} and provides a single argument to switch to the use of GPU accelerators, which can greatly enhance the computational scalibility of EMRI analysis. 

The currently available fast waveform models are the \texttt{FastSchwarzschildEccentricFlux} and \texttt{Pn5AAKWaveform}. The former is a fully relativistic waveform limited to the Schwarzschild eccentric regime. The latter combines a 5PN-integrated trajectory module from \cite{Fujita:2020zxe} and the semi-relativistic AAK waveform build methods from \cite{Chua2017} to produce a new AAK waveform that is available for generic Kerr inspirals, as well as more accurate and more robust than the original AAK waveform. Both waveforms can be accessed through a generic high-level waveform generator that provides a common interface to all waveform models generated in the source or detector frame. 

We then studied these waveforms to further understand the effect of harmonic mode content and their basic performance and characteristics when used in actual Bayesian posterior analysis. Leveraging GPU acceleration, we were able to move beyond Fisher matrix-type analysis and produce $\sim130$ full posterior distributions. The first main finding is that lower mode content chosen properly will lead to faster waveforms without sacrificing much accuracy in parameter estimation. However, this mode content must be relativistic and beyond just the quadrupolar mode, indicating the need for efficient mode selection tools. The second main takeaway is that using semi-relativistic amplitudes may strongly hinder the success of search and parameter estimation algorithms. Especially at higher MBH mass and higher eccentricity, strong biases and multimodal behavior are observed when injecting a relativistic waveform and attempting to extract that signal with a semi-relativistic template. With that said, a key point we have shown is the importance of phase overlap compared to amplitude overlap. The fact that a small number of modes that are relativistic and beyond the quadrupole can still provide successful tests shows the importance of matching phase over the duration of the inspiral. 

The FastEMRIWaveforms framework was originally built to bridge the gap between EMRI waveform modelling and data analysis. For the first time, we can generate fully relativistic waveforms at speeds fast enough for data analysis. While these waveforms are available in the Schwarzschild eccentric regime, the framework was designed to be flexible and adaptable to future exploration towards the generic Kerr background and post-adiabatic phase corrections. These new developments in EMRI data analysis with LISA provide an important step towards the goal of accomplishing the many forms of EMRI science within the LISA mission.  

\acknowledgments

MLK thanks Jonathan Gair and Ollie Burke for helpful discussions and Stas Babak for providing scripts for the conventions diagrams. AJKC acknowledges support from the NASA grant 18-LPS18-0027. NW acknowledges support from a Royal Society-Science Foundation Ireland Research Fellowship. This publication has emanated from research conducted with the financial support of Science Foundation Ireland under Grant number 16/RS-URF/3428. SAH's work on this problem was supported by NASA ATP Grant 80NSSC18K1091 and NSF Grant PHY-1707549. This research was supported in part through the computational resources and staff contributions provided for the Quest/Grail high performance computing facility at Northwestern University. This paper also employed use of \texttt{SciPy} \citep{scipy} and \texttt{Matplotlib} \citep{Matplotlib}. This work makes use of the Black Hole Perturbation Toolkit \cite{BHPToolkit}.

\providecommand{\noopsort}[1]{}\providecommand{\singleletter}[1]{#1}%

\end{document}